\documentclass[nonatbib]{elsarticle} 
\makeatletter 
\let\c@author\relax
\makeatother
\usepackage[size=normalsize,font=sf,labelfont=bf]{caption}
\usepackage{geometry}
\usepackage{subcaption}
\usepackage{graphicx}
\usepackage{amsthm}
\usepackage[colorlinks=true, pdfstartview=FitV, linkcolor=blue, citecolor=blue, urlcolor=blue]{hyperref}
\usepackage{threeparttable}
\usepackage{fancyhdr}
\usepackage{datetime} 
\usepackage{amsmath}
\usepackage{amssymb}
\usepackage{float}
\usepackage{sidecap} 
\usepackage[nodisplayskipstretch]{setspace} 

\usepackage{bm} 
\usepackage{makecell} 
\usepackage{dblfloatfix}    
\usepackage{multirow} 
\usepackage{booktabs} 
\usepackage{diagbox}
\usepackage{amssymb}
\usepackage[abbreviations]{foreign} 

\usepackage{textcomp} 
\usepackage[style=authoryear,backend=biber]{biblatex} 

\usepackage{xr-hyper} 
\usepackage{appendix}
\usepackage{siunitx}
\usepackage{derivative}

\usepackage[dvipsnames]{xcolor} 

\usepackage[draft]{pdfcomment} 
\usepackage[hyperfirst=false,acronym,sort=none,shortcuts,nopostdot,style=super,nonumberlist,toc,nogroupskip]{glossaries}

\glsdisablehyper 

\newcommand{\accolor}[1]{\textcolor{Sepia}{#1}}

\newacronymstyle{myacro}
{%
  \GlsUseAcrEntryDispStyle{long-short}%
}%
{%
  \GlsUseAcrStyleDefs{long-short}%
}

\setacronymstyle{myacro}

\newcommand*{\tip}[1]{
    \ifglsused{#1}{
      {\pdftooltip{\accolor{\glsentryshort{#1}}}{\glsentrydesc{#1}}}%
    }{%
      \gls{#1}
    }%
}%

\newcommand*{\tips}[1]{
    \ifglsused{#1}{
      {\pdftooltip{\accolor{\glsentryshortpl{#1}}}{\glsentrydescplural{#1}}}%
    }{%
      \glspl{#1}
    }%
}%

\newacronym{adc}{ADC}{Analog to Digital Converter}
\newacronym[description={A candidate or best-match Target Block for block matching in the t-d2 slice}]{tb}{TB}{Target Block}
\newacronym[description={A Reference Block centered on the event location in the t-d1 slice}]{rb}{RB}{Reference Block}
\newacronym[description={Coarse to Fine hierarchical search strategy used in BMOF}]{ctf}{CTF}{Coarse to Fine}
\newacronym[description={Coefficient Of Variation (sigma/mean)}]{cov}{COV}{Coefficient Of Variation}
\newacronym[description={Corner point in an image}]{kp}{KP}{Keypoint}
\newacronym[description={Direction Selective model of motion detection in biological vision, usually either Hassenstein-Reichhardt or Barlow-Levick type}]{ds}{DS}{Direction Selective}
\newacronym[description={DVS optical flow method that fits a plane to the local event cloud}]{lp}{LP}{Local Plane}
\newacronym[description={False Negative Rate; signal that is incorrectly classified as noise}]{fnr}{FNR}{False Negative Rate}
\newacronym[description={False Positive Rate; noise that is incorrectly classified as signal}]{fpr}{FPR}{False Positive Rate}
\newacronym[description={Inter Spike Interval (nomenclature from neuroscience)}]{isi}{ISI}{Inter Spike Interval}
\newacronym[description={Multipurpose block random access memory module in FPGA}]{bram}{BRAM}{Block RAM}
\newacronym[description={Register Transfer Logic intermediate form, consisting of combinational and synchronous register logic cells}]{rtl}{RTL}{Register Transfer Logic}
\newacronym[description={Search Area for block matching}]{sa}{SA}{Search Area}
\newacronym[description={Single Threshold Metric; a measure of the ROC TPR/FPR tradeoff at one discrimination threshold}]{stm}{STM}{Single Threshold Metric}
\newacronym[description={Slice-based FAST that uses accumulated event count slices for detecting keypoints}]{sfast}{SFAST}{Slice-based FAST}
\newacronym[description={Surface of Active Event; image of latest event timestamps, same as Timestamp Image}]{sae}{SAE}{Surface of Active Events}
\newacronym[description={System on Chip; FPGA with embedded programmable processor}]{soc}{SoC}{System on Chip}
\newacronym[description={Timestamp Image; image of latest event timstamps, same as Surface of Active Events}]{ti}{TI}{Timestamp Image}
\newacronym[description={True Negative Rate; noise that is correctly classified as noise}]{tnr}{TNR}{True Negative Rate}
\newacronym[description={True Positive Rate; signal that is correctly classified as signal}]{tpr}{TPR}{True Positive Rate}
\newacronym[description={Visual Odometry}]{vod}{VOD}{Visual Odometry}
\newacronym[longplural={First In First Out memories}]{fifo}{FIFO}{First In First Out memory}
\newacronym{aae}{AAE}{Average Angular Error}
\newacronym{slowmo}{SlowMo}{slow motion}
\newacronym{abmof}{ABMOF}{Adaptive Block Matching Optical Flow}
\newacronym{aee}{AEE}{Average Endpoint Error}
\newacronym{aer}{AER}{Address Event Protocol}
\newacronym{aop}{AoP}{Angle of Polarization}
\newacronym{aps}{APS}{Active Pixel Sensor}
\newacronym{asic}{ASIC}{Application Specific Integrated Circuit}
\newacronym{auc}{AUC}{Area Under the Curve}
\newacronym{baf}{BAF}{Background Activity Filter}
\newacronym{ba}{BA}{Background Activity}
\newacronym{bmof}{BMOF}{Block Matching Optical Flow}
\newacronym{bm}{BM}{Block Matching}
\newacronym{cfa}{CFA}{Color Filter Array}
\newacronym{cf}{CF}{Complementary Filter}
\newacronym{cis}{CIS}{CMOS Image Sensor}
\newacronym{cnn}{CNN}{Convolutional Neural Network}
\newacronym{cots}{COTS}{Commodity Off-The-Shelf}
\newacronym{cpu}{CPU}{Central Processing Unit}
\newacronym{cv}{CV}{Computer Vision}
\newacronym{davis}{DAVIS}{Dynamic and Active Pixel Vision Sensor}
\newacronym{dnn}{DNN}{Deep Neural Network}
\newacronym{dop}{DoLP}{Degree of Linear Polarization}
\newacronym{dr}{DR}{Dynamic Range}
\newacronym{dram}{DRAM}{Dynamic RAM}
\newacronym{drcn}{DRCN}{Deep Recurrent Convolutional Network}
\newacronym{dsp}{DSP}{Digital Signal Processing unit}
\newacronym{dvs}{DVS}{Dynamic Vision Sensor}
\newacronym{dvsd}{DVSD}{Dynamic Vision Sensor Disdrometer}
\newacronym{dwf}{DWF}{Double Window Filter}
\newacronym{edp}{EDP}{Event Denoising Precision}
\newacronym[description={Extinction Ratio (the reciprocal of the ratio of cross-polarized light that passes the linear polarizer}]{er}{ER}{Extinction Ratio}
\newacronym{fom}{FOM}{Figure of Merit}
\newacronym{fov}{FoV}{field of view}
\newacronym{fpga}{FPGA}{Field Programmable Gate Array}
\newacronym{fpn}{FPN}{Fixed Pattern Noise}
\newacronym{fps}{FPS}{frames per second}
\newacronym{fsae}{FSAE}{Filtered Surface of Active Events}
\newacronym{fwf}{FWF}{Fixed Window Filter}
\newacronym{gpu}{GPU}{Graphics Processing Unit}
\newacronym{gt}{GT}{ground truth}
\newacronym{hddg}{HDDG}{Hard Disk Droplet Generator}
\newacronym{hdd}{HDD}{hard disk drive}
\newacronym{hdl}{HDL}{Hardware Description Language}
\newacronym{hdr}{HDR}{high dynamic range}
\newacronym{hls}{HLS}{High Level Synthesis}
\newacronym{icm}{ICM}{Iterated Conditional Modes}
\newacronym{id}{ID}{Index Decay}
\newacronym{iir}{IIR}{Infinite Impulse Response}
\newacronym{inceptiveevent}{IE}{Inceptive Event}
\newacronym{iot}{IoT}{Internet of Things}
\newacronym{ip}{IP}{Intellectual Property}
\newacronym{its}{ITS}{Invariant Time Surface}
\newacronym{ivdg}{IVDG}{Intravenous Dripper Droplet Generator}
\newacronym{iv}{IV}{intravenous}
\newacronym{jwd}{JWD}{Joss–Waldvogel Disdrometer}
\newacronym{li}{LI}{Leaky Integrator}
\newacronym{lk}{LK}{Lucas-Kanade}
\newacronym{mpeg}{MPEG}{Motion Picture Experts Group}
\newacronym{na}{NA}{Numerical Aperture}
\newacronym{nir}{NIR}{Near Infrared}
\newacronym{nnb}{NNb}{Nearest Neighbor}
\newacronym{of}{OF}{Optical Flow}
\newacronym{onf}{ONF}{Order(N) Filter}
\newacronym{parsivel}{PARSIVEL}{Particle Size Velocity}
\newacronym{pcb}{PCB}{Printed Circuit Board}
\newacronym{pd}{PD}{photodiode}
\newacronym{pdavis}{PDAVIS}{Polarization Dynamic and Active Pixel Vision Sensor}
\newacronym{pfa}{PFA}{Polarization Filter Array}
\newacronym{pl}{PL}{programmable Logic}
\newacronym{por}{POR}{Positive Output Ratio}
\newacronym{prm}{PRM}{Pixel Rendering Module}
\newacronym{ps}{PS}{Processing System}
\newacronym{pugm}{PUGM}{Probabilistic Undirected Graph Model}
\newacronym{qwp}{QWP}{Quarter Wave Plate}
\newacronym{ram}{RAM}{Random Access Memory}
\newacronym{ratp}{RATP}{Recursive Adaptive Temporal Pooling}
\newacronym{roc}{ROC}{Receiver Operating Characteristic}
\newacronym{roi}{ROI}{Region of Interest}
\newacronym{rpmd}{RPMD}{Relative Plausibility Measure of Denoising}
\newacronym{rpm}{RPM}{Revolutions per Minute}
\newacronym{sad}{SAD}{Sum of Absolute Differences}
\newacronym{sd}{SD}{Secure Digital}
\newacronym{silc}{SILC}{Speed Invariant Learned Corners}
\newacronym{sits}{SITS}{Speed Invariant Time Surface}
\newacronym{slam}{SLAM}{Simultaneous Localization And Mapping}
\newacronym{sm}{SM}{Supplementary Material}
\newacronym{snr}{SNR}{Signal to Noise Ratio}
\newacronym{soa}{SoA}{State of the Art}
\newacronym{sram}{SRAM}{Static RAM}
\newacronym{stcf}{STCF}{SpatioTemporal Correlation Filter}
\newacronym{susan}{SUSAN}{Smallest Univalue Segment Assimilating Nucleus)}
\newacronym{tda}{TDA}{Time Decay Adapted}
\newacronym{td}{TD}{time decay}
\newacronym{timsl}{TS}{time slice}
\newacronym{usb}{USB}{Universal Serial Bus}
\newacronym{vga}{VGA}{Video Graphics Adaptor}
\newacronym{vhdl}{VHDL}{Very High-Speed Integrated Circuit Hardware Description Language}
\newacronym{zoh}{ZOH}{Zero-Order Hold}
\newacronym{dsd}{DSD}{Drop Size Distribution}
\newacronym{dvd}{DVD}{Drop Velocity Distribution}
\newacronym{mape}{MAPE}{Mean Absolute Percentage Error}
\newacronym{ape}{APE}{Absolute Percentage Error}
\newacronym{mae}{MAE}{Mean Absolute Error}
\newacronym{std}{STD}{Standard Deviation}
\newacronym{rmsd}{RMSD}{Root-Mean-Square Deviation}
\newacronym{2dvd}{2DVD}{2-Dimensional Video Disdrometer}
\newacronym{psvd}{PSVD}{Particle Size and Velocity Distribution}
\newacronym{dof}{DoF}{Depth of Field}
\newacronym{aov}{AoV}{Angle of View}
\newacronym{led}{LED}{Light Emitting Diode}
\newacronym{los}{LoS}{Line of Sight}
\newacronym{pof}{PoF}{Plane of Focus}
\newacronym{pofr}{PoFR}{Plane of Focus Rectangle}
\newacronym{imu}{IMU}{Inertial Measurement Unit}

\DeclareSourcemap{
  \maps[datatype=bibtex]{
    \map[overwrite]{
      \step[fieldsource=doi, final]
      \step[fieldset=url, null]
      \step[fieldset=eprint, null]
    }  
  }
}

\AtEveryBibitem{%
  \clearlist{language}
}
\makeatother

\makeatletter
\newcommand*{\addFileDependency}[1]{
  \typeout{(#1)}
  \@addtofilelist{#1}
  \IfFileExists{#1}{}{\typeout{No file #1.}}
}
\makeatother

\newcommand{\MATLAB}{\textsc{Matlab}\xspace}

\begin{document}

\begin{frontmatter}

\title{Measuring diameters and velocities of artificial raindrops with a neuromorphic dynamic vision sensor disdrometer}

\author[1]{Jan Steiner \fnref{fn1}} 
\author[1]{Kire Micev \fnref{fn1}}
\author[2]{Asude Aydin}
\author[3]{Jörg Rieckermann}
\author[2]{Tobi Delbruck\corref{cor1}}
\ead{tobi@ini.uzh.ch}

\cortext[cor1]{Corresponding author}
\address[1]{Department of Mechanical and Process Engineering, ETH Zurich, Zurich, Switzerland}
\address[2]{Institute of Neuroinformatics, University of Zurich and ETH Zurich, Zurich, Switzerland}
\address[3]{EAWAG, Swiss Federal Institute of Aquatic Science and Technology, Dübendorf, Switzerland}

\fntext[fn1]{These authors contributed equally.}

\begin{abstract}
Hydrometers that can measure size and velocity distributions of precipitation are needed for research and corrections of rainfall estimates from weather radars and microwave links. Existing video disdrometers measure drop size distributions, but underestimate small raindrops and are impractical for widespread always-on IoT deployment. We propose an innovative method of measuring droplet size and velocity using a neuromorphic event camera. These dynamic vision sensors asynchronously output a sparse stream of pixel brightness changes. Droplets falling through the plane of focus create events generated by the motion of the droplet. Droplet size and speed are inferred from the stream of events. Using an improved hard disk arm actuator to reliably generate artificial raindrops, our experiments show small errors of 7\% (maximum mean absolute percentage error) for droplet sizes from 0.3 to 2.5\,mm and speeds from 1.3\,m/s to 8.0\,m/s. Each droplet requires the processing of only a few hundred to thousands of events, potentially enabling low-power always-on disdrometers that consume power proportional to the rainfall rate. 
\end{abstract}


\end{frontmatter}

\section{Introduction}  \label{sec:intro}


There are increasing numbers of optical disdrometers that measure the diameter and speed of hydrometeors at ground level \autocite{liu2013comparison,johannsen_comparison_2020}. Their \tip{dsd} measurements can be combined with weather radars or microwave links to predict a \tip{dsd} over a larger area \autocite{kruger_two-dimensional_2002,spackova_year_2021}.
The \tip{soa} scientific instrument is the \tip{2dvd} first described by \textcite{kruger_two-dimensional_2002}\footnote{See also \href{https://www.distrometer.at/}{www.distrometer.at}}.  
However, \tip{2dvd} and competing \tip{parsivel} laser-sheet disdrometers have been reported to underestimate total rainfall volume and drift over time resulting in unpractical long-term deployment \autocite{johannsen_comparison_2020,jaffrain_experimental_2011,upton_investigation_2008}. Different types of disdrometers have been shown to produce measured \tip{dsd}s that differ dramatically for small droplets \autocite{johannsen_comparison_2020, cao_analysis_2008}. They are too expensive for ubiquitous deployment, and consume a lot of power on the order of 100\,W making them impractical for solar-powered weather monitoring, where brownouts can occur in dark weather conditions~\autocite{spackova_year_2021}. Therefore, the ideal disdrometer would be precise and low-cost and would enable autonomous continuous \tip{dsd} measurements by using less power when there are fewer droplets to measure.


In this paper, we propose using a novel droplet-driven sampling approach based on analyzing the brightness change events produce by a \tip{dvs} event camera. Such an event camera does not capture stroboscopic images using a shutter as a conventional camera. Instead, each pixel reports asynchronous changes in brightness as they occur, and stays silent otherwise (Fig.~\ref{fig:fig1_methods}A)~\autocite{Lichtsteiner2008-dvs,Gallego2020-survey-paper}. They have been successfully used in many high speed robotics and machine vision applications \parencite{Gallego2020-survey-paper}, but not yet in environmental or atmospheric monitoring.

Our main contributions are:
\begin{enumerate}
    \item We propose a novel optical disdrometer method that exploits the activity-driven output and high time resolution of \tip{dvs} brightness change events to efficiently measure individual droplet size and speed
    using the shallow \tip{dof} of a fast lens to localize individual droplets in 3d space.
    \item We generate high-quality ground-truth data for the droplets by modifying the \tip{hddg} from \textcite{Kosch2015-hdd-droplet-generator} and report how to reproduce this \tip{hddg}.
    \item We report the first measurements of droplet size and speed with our proposed \tip{dvsd} and show that the \tip{dsd} satisfactorily aligns with the ground truth data with at most a mean absolute percentage error of 7\%.
\end{enumerate}


\section{Materials and Methods}
\label{sec:methods}

\subsection{\tip{dvsd} Setup}
\label{sec:methods_dvsd}
\begin{figure}[h!]
    \centering
    \includegraphics[width=.9\textwidth]{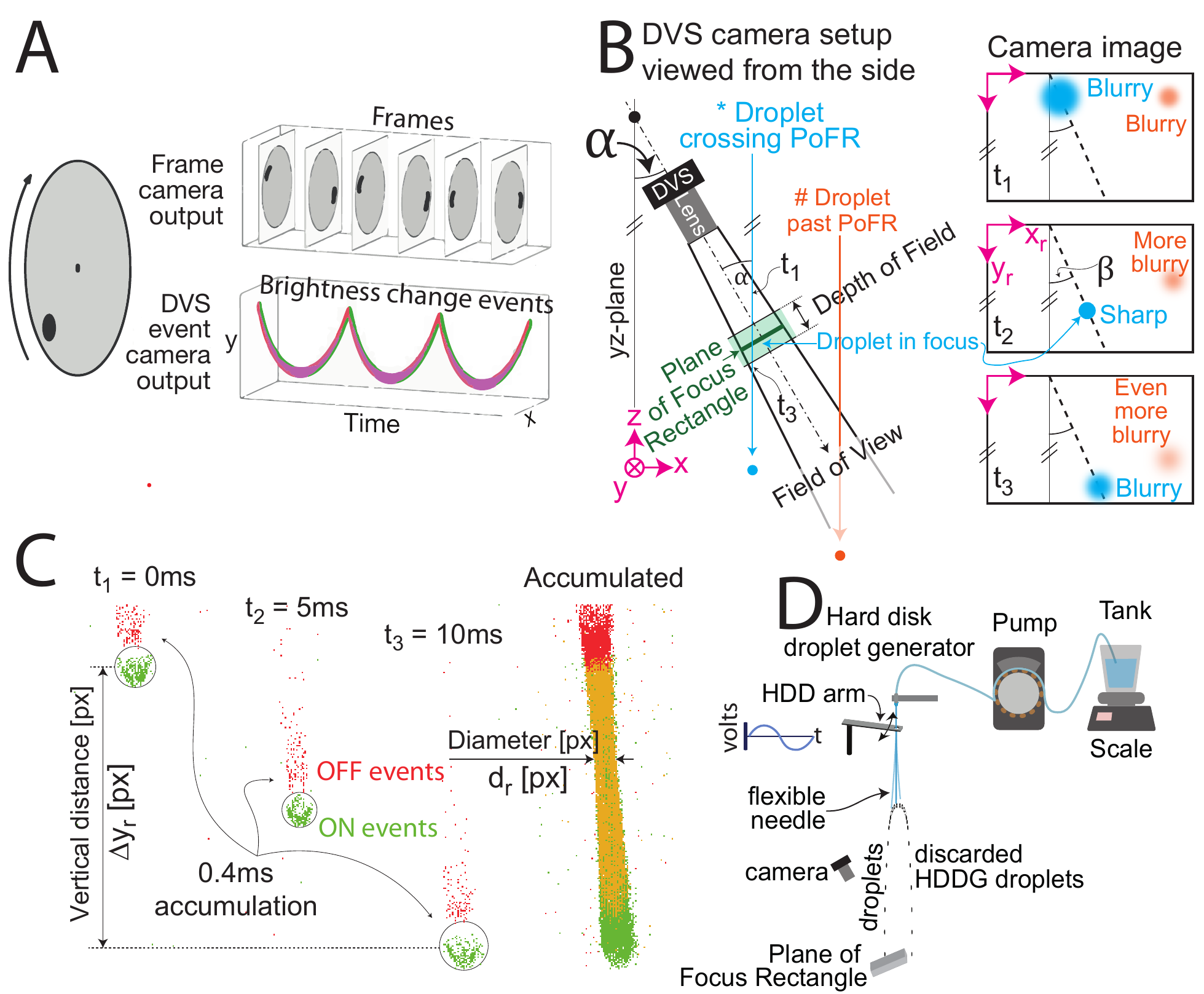}
    \caption{\textbf{\glsreset{dvsd} Summary of \tip{dvsd} methods.} 
    \textbf{A}: Comparison between a conventional frame camera and a \tip{dvs} capturing a rotating disk with a black dot. The frame camera outputs frames with finite exposure duration at discrete time intervals, whereas the event camera continuously outputs brightness change events, which results in a helix of discrete events in the space time plot (green: increase in brightness, red: decrease in brightness) \parencite{Lichtsteiner2008-dvs,Gallego2020-survey-paper} (\tip{sm} \ref{subs:dvs_camera}). 
    \textbf{B}: Side view of the \tip{dvs} camera setup in experiments and three illustrations of \tip{dvs} recordings. 
    The cyan droplet enters and exits the \tip{fov}, which is tilted at a small angle $\alpha$ from the vertical $yz$-plane; 
    we used 22$\textdegree$ for \tip{hddg} and 29$\textdegree$ for \tip{ivdg} experiments. The corresponding recording is illustrated on the right side at three different times: 
    the cyan droplet entering the \tip{fov}, droplet crossing the \tip{pof}, and droplet exiting the \tip{fov}. 
    $\beta$ is the angle of the droplet from the vertical $y_\text{r}$-axis seen on the recording, caused by droplet velocity component in the $yz$-plane (out of the page) resulting from the \tip{hddg}. 
    The orange droplet never crosses the \tip{pof} and only grows increasingly blurry. (\tip{sm} \ref{subs:experimental_setup},\ref{subs:experiments})
    \textbf{C}: Sample \tip{dvs} recording of a droplet crossing the \tip{pof}, which is demonstrated in three frames with 5\,ms time differences between each frame. 
    Each of the three \tip{dvs} frames in this sample is an accumulation of 0.4\,ms of events. Green points correspond to ON events, red points show OFF,  and yellow points show overlapping of ON and OFF events. 
    The rightmost frame shows all accumulated events over 10\,ms. 
    Each droplet creates several hundred to several thousand events, depending on its size. 
    We estimate the falling speed $v_\text{r}$ by measuring the focal plane speed of the droplet. 
    The diameter $d_\text{r}$ of the droplet is measured at the waist of the hourglass when the droplet is in focus, 
    as illustrated on the right. 
    Eqs.~\eqref{eq:d_DVS} and \eqref{eq:v_DVS} provide the droplet diameter and speed.
    (\tip{sm} \ref{subs:jaer_measurement})
    \textbf{D}: \tip{hddg} modified from \textcite{Kosch2015-hdd-droplet-generator}. 
    The droplet generator uses a hard disk actuator to oscillate a flexible needle with a constant flow rate of water fed into the needle from a pump. 
    The water tank is placed on top of a scale to calculate the flow rate. Within a 4$\times$ range of oscillation frequencies, 
    a droplet is released at each end of the oscillation by large acceleration forces acting on the flexible needle. 
    The diameter of droplets released from the needle is adjusted by the oscillation frequency of the needle. 
    We generated the large 2.5\,mm droplets falling 10\,m through a circular staircase well with an \tip{iv} dripper. (\tip{sm} \ref{subs:drop_creation} \ref{subs:experimental_setup})}
    \label{fig:fig1}
    \label{fig:fig1_methods}
\end{figure}

Fig.~\ref{fig:fig1_methods} illustrates our proposed \tip{dvsd} method, which is detailed in our \tip{sm} Secs.~\ref{subs:drop_creation} and \ref{subs:experimental_setup}. 
The \tip{dvs} camera (Fig.~\ref{fig:fig1_methods}A, \tip{sm} Sec.~\ref{subs:dvs_camera}) asynchronously reports brightness change events as the droplets pass through a thin \tip{dof} at the \tip{pofr} from a lens that looks down on the rainfall from a steep angle (Fig.~\ref{fig:fig1_methods}B). Each droplet produces a few hundred to a few thousand \tip{dvs} brightness change events. 
By a simple analysis of this cluster of events, the \tip{dvsd} can measure both the size and the speed of the droplet.
We developed a modified \tip{hddg} to generate small droplets and used an \tip{ivdg} for large droplets. 
Fig. \ref{fig:fig1}C shows an illuminated falling water droplet recorded with the \tip{dvs} camera. 
Our method consists of two key principles. First, we aim the camera downward at a steep angle, 
with an angle $\alpha$ from the vertical (Fig. \ref{fig:fig1}B: left). 
Second, the diameters of the droplets crossing the shallow \tip{dof} at the \tip{pof} are measured unambiguously, 
\ie, since the \tip{pof} is located at a fixed working distance from the lens, 
we can infer the 3d position of the droplet, and hence disambiguate the absolute size from the image size. Droplets passing through the camera's \tip{pof} come into focus, showing a high contrast, whereas droplets outside the \tip{pof} appear blurry.
Therefore, droplets that are out of focus cover a larger area of the recording than when they are in focus (see Fig. \ref{fig:fig1}C: black circles). 
Accumulating the events belonging to one droplet that crosses the \tip{pofr} (marked by * in figure) produces an hourglass shape (Fig.~\ref{fig:fig1}C: accumulated events) where the ideal moment for a droplet diameter measurement (Sec.~\ref{subs:jaer_measurement}) is at the waist of this hourglass. The hourglass should be as concave as possible to facilitate the detection of the waist. Using a fast lens with a small aperture ratio $f$ number produces a shallow \tip{dof}, increasing the amount of blur of the droplets that are out of focus. 
Fig. \ref{fig:fig1}B also illustrates how a droplet that crosses the \tip{fov} but past the \tip{pofr} (marked by \# in figure) creates an accumulated image that starts out blurry and becomes increasingly blurry until it leaves the \tip{fov}; similarly (but not illustrated), a droplet that crosses the \tip{fov} in front of \tip{pofr} creates an accumulated event image that starts out blurry and becomes increasingly sharper until it leaves the \tip{fov}. 

\subsection{Modification of the \tip{hddg}}
\label{methids_hddg}

Fig.~\ref{fig:fig1}D and \tip{sm} Sec.~\ref{subs:drop_creation} and Figs.~\ref{fig:hddg_sketch} and \ref{fig:experimental_setup_HDD} illustrate our \tip{hddg}. It is based on previous work by \textcite{Kosch2015-hdd-droplet-generator}, who utilized a computer hard disk arm as an actuator.
They used a high-frequency buzz to create ripples in a steady stream of water emitted by a stiff glass needle, which would break up into small droplets. Our \tip{hddg} uses a flexible plastic needle, which, if properly combined with a steady stream of water, creates a single droplet at each end of an oscillation, resulting in two droplet streams, one of which we measured. 
We used a discarded hard disk drive that we disassembled to expose the platter head actuator arm.
The arm is coupled to the needle by threading the needle through adhesive tape applied over the hole in the arm allowing the needle to protrude. The arm is actuated with a home audio power amplifier driven by sinusoidal waveforms generated by an audio wave generating program where we used coil driver amplitudes from
\SIrange{5}{20}{Vpp} 
and frequencies from 
\SIrange{60}{220}{Hz}. 
\section{Results} \label{sec:results}

\begin{figure}[h!]
    \centering
    \includegraphics[width=\textwidth]{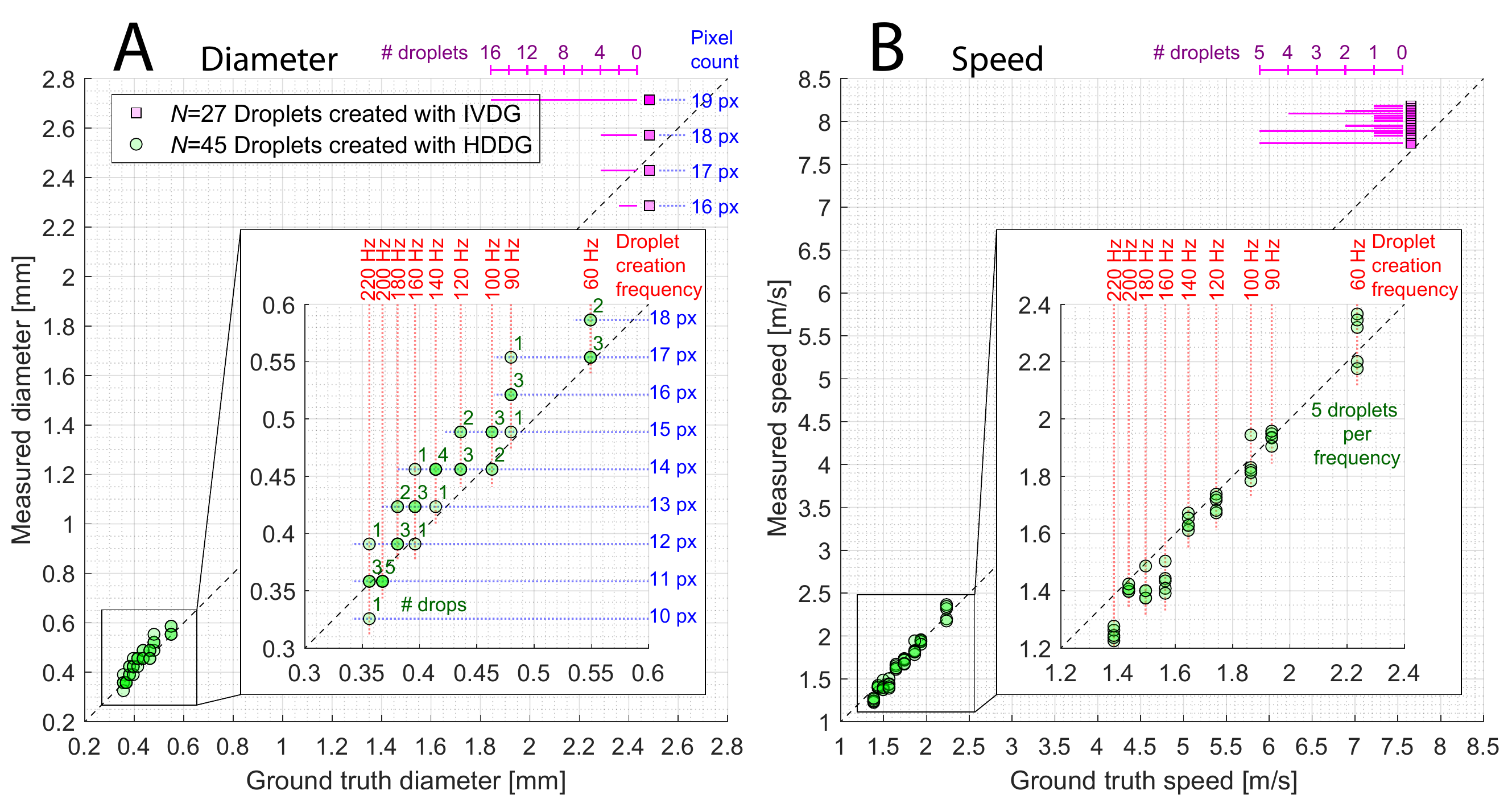}
    \caption{\textbf{Main results.} \tip{dvsd} measurement results of droplets compared to \tip{gt} values, where \textbf{A} shows the diameter and \textbf{B} shows the velocity. The dashed black line represents a 45°-line passing through the origin. Both drop creation methods are included in the plots. The zoomed plots show droplets created with the \tip{hddg} for improved visibility. Numbers adjacent to the points on the left zoomed plot indicate overlaps whereas the zoomed plot on the right has 5 droplets per frequency. The \tip{ivdg} droplets are shown in purple with a histogram for number of overlapping points. Quantization effects caused by the pixel count or frequency are illustrated as a grid pattern in the plots where the effect is significant.  (See \tip{sm} \ref{subs:experiments} for details.)}
    \label{fig:fig2_results}
\end{figure}

We conducted two series of experiments, one with the \tip{hddg} and one with the \tip{ivdg} (\tip{sm} \ref{subs:drop_creation}). 
We used different lenses to make it easier to capture droplets crossing the \tip{pof}. 
The droplets created by the \tip{hddg} ranged from 0.3\,mm to 0.6\,mm (10 to 20 pixel diameter on the image), while the droplets created by the \tip{ivdg} were 2.5\,mm (17-18 pixel diameter). In both experiments, the height of the fall was sufficient for the droplets to reach within 97\% of the terminal speed (\tip{sm} \ref{subs:simulation}). Fig. \ref{fig:fig2_results} compares the measurement results performed with the \tip{dvs} (see \ref{subs:jaer_measurement}) to \tip{gt} (see \tip{sm} \ref{subs:ground_truth} and \ref{subs:simulation}). 
The \tip{hddg} droplets (green data points) are magnified for better visibility, and the \tip{ivdg} droplets (purple data points) have a purple histogram beside them, indicating the number of measurement results that overlap. 

The results show excellent linearity over the entire measurement range for both size and speed; the dashed line in each plot has a slope of one and passes through the origin; it lies close to both small and large droplet measurements. Size measurements slightly overestimate small droplet diameters, and speed measurements slightly underestimate large droplet speeds. The quantization of the data arises from the quantized droplet size generation and the pixel discretization. Horizontal quantization is caused by the quantized \tip{hddg} droplet creation frequencies, which control the diameters of the droplets. Vertical quantization is caused by the low pixel count of the diameter of the droplets in the \tip{dvs} recording. The speed measurements do not have any significant vertical quantization effects, due to large pixel displacements ($\approx$ 100 pixels) and the fine \tip{dvs} event timestamp resolution of 1\,µs. 

 We used \tip{mape} to further quantify the discrepancy between the ground truth values and the \tip{dvs} measurements (\tip{sm} \ref{subs:error_analysis}). Table \ref{tab:mape} lists the diameter and velocity \tip{mape} for both experiments. In all cases, \tip{mape} remains below 7\%.
\begin{SCtable}[10][h!]
    \centering
    \renewcommand{\arraystretch}{1.2}
    \begin{tabular}{|l | c | c|} 
        \multicolumn{3}{c}{\textbf{Mean absolute percentage error}} \\
        \hline
        Experiment  & Diameter & Velocity \\
        \hline\hline
        HDDG            & 6\%      & 4\%      \\
        IVDG            & 7\%      & 4\%      \\
        \hline
    \end{tabular}
    \caption{\tip{mape} of the \tip{dvs} measurements compared to \tip{gt} values, for the diameter and velocity measurements from both experiments.}
    \label{tab:mape}
\end{SCtable}

Table \ref{tab:uncert} lists the estimated combined uncertainties of the measured diameter and velocity using the method explained in \tip{sm} \ref{subs:error_analysis}. 
 In some cases, a range of uncertainties is given, which means the uncertainty depends on the size of the droplet. The combined uncertainty percentage is largest for the \tip{dvs} measurement of the smallest droplets created with the \tip{hddg}, 
which had a diameter of $0.35\,\text{mm}\pm0.03\,\text{mm}$ ($\pm 10\%$). 
This large uncertainty is caused by the low pixel diameter count ($\approx$ 10 pixels).

The uncertainty (\ie precision) of \tip{dvsd} diameter and velocity measurements are mainly limited by the spatial resolution of the $346\times 260$-pixel \tip{dvs} (see Table \ref{tab:uncert}: bottom). 
The \tip{gt} diameter uncertainty (see Table \ref{tab:uncert}: top) was mostly caused by the measurement uncertainty of the scale and noise in the droplet generation by the \tip{hddg} and \tip{ivdg}.
The \tip{gt} velocity uncertainty arises from neglecting air turbulence, droplet deformation, and inaccurate sphere model values, i.e. \tip{gt} droplet diameter estimates from mass flow. 
The \tip{gt} droplet diameter and velocity are calculated from their mass and from the simulation, respectively (see \ref{subs:drop_creation} and \ref{subs:simulation}).
Therefore, if the diameter is uncertain, it increases velocity uncertainty.

\begin{SCtable}[10][ht!]
    \centering
    \renewcommand{\arraystretch}{1.2}
    \begin{tabular}{| l l | c c | c c |} 
        \multicolumn{6}{c}{\textbf{Combined uncertainty}}    \\
        \hline
        \multirow{2}{*}{Method}  & \multirow{2}{*}{Experiment} & \multicolumn{2}{c|}{Diameter}  & \multicolumn{2}{c|}{Velocity} \\
             &    & $\pm$[\%]       & $\pm$[mm]     & $\pm$[\%]      & $\pm$[m/s]    \\
        \hline\hline
        \multirow{2}{*}{\tip{gt}}   & \tip{hddg} & 3  & \numrange{0.01}{0.02}  & \numrange{4}{7}  &  0.1   \\
                                        & \tip{ivdg}   &  2 & 0.05      &  4    &  0.3         \\
        \hline
        \multirow{2}{*}{\tip{dvs}}& \tip{hddg}   &  \numrange{6}{10}   &  0.03 &  7    &  \numrange{0.1}{0.2} \\
                             & \tip{ivdg}   &  \numrange{6}{7}    &  0.15      &  6    &  0.5     \\
        \hline
    \end{tabular}
    \caption{Percentage and absolute combined uncertainty of the \tip{dvs} measurements and \tip{gt} values for diameter and velocity.}
    \label{tab:uncert}
\end{SCtable}
\section{Discussion} \label{sec:discussions}

\subsection{Experimental results}
Although the large droplets were generated and measured differently than the small droplets, 
the two data sets are very consistent (see Fig. \ref{fig:fig2_results}: green and purple data sets). 
The offset of the data points from the 45°-line (see Fig. \ref{fig:fig2_results}) 
is correctable since it arises from slightly inaccurate $\alpha$ and $M$ estimates. 
Therefore, we believe that the \tip{dvsd} can achieve accurate droplet measurements.

We used a shorter lens for the larger \tip{ivdg} droplets only to allow us to capture the large droplets more easily, 
since they scattered much more from random wind currents in the staircase than the small droplets from the \tip{hddg}. It is possible to obtain better precision of large droplets by using the same lens, but with the trade-off of longer experiment time since fewer will pass through the \tip{pofr}.

Future studies should compare the \tip{dvsd} directly to \tip{soa} disdrometers that measure individual droplet diameters and velocities. 
Improving our \tip{hddg} should be investigated since our \tip{hddg} was somewhat unstable, which required patience to capture sufficient good droplets to measure (\tip{sm} \ref{subs:stability_of_hddg}).
Using \tip{nir} illumination should also be tried, since most insects would be blind to it and hence not be attracted to the \tip{dvsd}, and \tip{dvs} silicon photodiodes work well with \tip{nir} illumination.

\subsection{Limitations of experiments}

Our experiments were carried out in a controlled environment using two droplet generators, i.e. \tip{hddg} and \tip{ivdg}. 
However, unlike real rainfall conditions, there were no strong winds. 
Moreover, the drop jets were localized and did not occlude each other. 
We do not believe that occlusion would be a problem due to the optical arrangement,
but the droplet tracks could merge or overlap and the droplets in front or behind the \tip{pof} could disturb the measurements. 
Therefore, it is difficult to predict how well a \tip{dvsd} would perform under windy conditions or with heavy rainfall. 

Our \textit{hourglass} \tip{dvsd} method (\tip{sm} \ref{subs:jaer_measurement}) works best when the droplets pass all the way through the \tip{fov} (see Fig.~\ref{fig:fig1}B: left, and Fig.~\ref{fig:setup}: bottom left corner). 
In an extreme case, the wind could make a droplet trajectory parallel to the \tip{los} of the camera. 
If this is the case, no hourglass would be visible on the \tip{dvs} recording after an accumulation of events; from the point of view of the \tip{dvs}, 
the droplets would appear to shrink and grow while slowly drifting in a random direction. 
In principle, it should be possible to infer the 3d trajectory of the droplet by developing an algorithm that continuously estimates the diameter and velocity of the droplets.
We would base such algorithm on cluster trackers commonly used for other \tip{dvs} applications \autocite{delbruck2013robotic,Gallego2020-survey-paper}.
These trackers would initiate clusters at the top of the \tip{fov}, and then use brightness change events to track the droplets, 
while measuring the droplet velocity and diameter. 
A simple set of plausibility checks on the cluster path and a fit to the hourglass diameter samples could provide the image plane droplet measurements along with their uncertainties.

The size of the sampling area plays an important role in how quickly a \tip{dsd} can be obtained. 
The sampling area of the \tip{dvsd} decreases slightly with increasing drop size, because the droplets must be fully inside and pass through the \tip{fov}. 
Therefore, a correction will be needed to estimate the \tip{dsd} to account for the smaller fraction of larger droplets that are measured.

If our \tip{dvs} were required to have the same sampling area as the OTT Parsivel$^2$ (see Table \ref{tab:specs}), a reduction in focal length would be necessary, 
but would increase the current 0.35\,mm drop size uncertainty from 10\% (see Table \ref{tab:uncert}) to 75\%. 
However, these limitations are a result of the low spatial resolution of our prototype camera and  megapixel \tip{dvs} are already available~\autocite{suh20201280}. 
With that \tip{dvs}, the 0.35\,mm droplet size uncertainty would be about 20\% 
while matching the sampling area of the OTT Parsivel$^2$ by adjusting the focal length to the appropriate value.

\subsection{Comparing \tip{dvsd} to other optical disdrometers}
Table \ref{tab:specs} compares the specifications of our current \tip{dvs} prototype to the OTT Parsivel$^2$ and \tip{2dvd}. 
The \tip{dvsd} takes advantage of the ability of the \tip{dvs} to finely measure the velocity of the droplet across the plane of focus 
and uses the \tip{pof} to locate the droplet in space for unambiguous size measurement. 
Other optical disdrometers measure the size of the droplets by the size of the 1D occlusion (\tip{2dvd}) or the decrease in the intensity of light (\tip{parsivel}).

The sampling uncertainty of our disdrometer is in a range similar to that of other optical devices. Field experiments with co-located instruments resulted in about 5\%–11\% error in small drops (D0 = 1–2\,mm) and the error varies from approximately 8\% to 4.5\% at D0 = 1.5\,mm from 1 min to 10 min sampling time~\autocite{Jaffrain2012-network-of-parsivel,Chang2020-uncertainty-four-colocated}.

\textcite{johannsen_comparison_2020} reported difficulty with long-term measurements using \tip{parsivel} and \tip{2dvd} because of drift and  insect and spider debris accumulating in optical housings. The simpler free-space optical arrangement of the \tip{dvsd} could be advantageous in avoiding these problems. Speed is measured by the time of passage between nearby light sheets (\tip{2dvd}) or by the time that a single light sheet is occluded (\tip{parsivel})~\parencite{johannsen_comparison_2020}. Both techniques require high sample rates because droplets that fall at a terminal speed pass through any given point in a few hundred microseconds. 
\Eg a 1\,mm droplet falling at its terminal speed of 4\,m/s (\tip{sm} \ref{subs:simulation}) passes by in only 250\,$\mu$s. 
The 1\,$\mu$s time resolution of \tip{dvs} allows very accurate measurements of droplet speed in the image plane, but at a low camera data rate of a few hundred to thousand brightness change events per droplet, which could easily be processed by an embedded microcontroller.

\begin{table}[t] 
    \begin{threeparttable}[h]
    \caption{\textbf{Comparison of disdrometer specifications.} Data may not be accurate for latest models.}
         \label{tab:specs}

    \begin{tabular}{r|ccc}
        \multirow{2}{*}{Specification}       & \multicolumn{3}{c}{Device}                     \\
                            &  \tip{dvsd}\tnote{1}      & \tip{2dvd}\tnote{2}     & \tip{parsivel}\tnote{3}    \\ 
        \hline
        Technology       & 1 dynamic vision sensor   & 2 line-scan cameras   & Laser-sheet  \\ 
        Sensor resolution  & $346\times 260$   & 512\,px   & 1 photodetector  \\ 
        Pixel pitch & 18.5$\mu$m & NA & none \\
        Power               & 3\,W camera + 40\,W LED    & 500\,W        & 100\,W     \\ 
        Data rate           & variable (0-1MB/s)              & 80\,MB/s             & 2.4\,MB/s          \\ 
        Optics & 300mm (\tip{hddg}) 75mm (\tip{ivdg}) & NA & NA\\
        Sampling area       & 0.88\,cm$^2$ (\tip{hddg})  400mm\,cm$^2$ (\tip{ivdg})      & 100 cm$^2$          & 54\,cm$^2$          \\
        Diameter range    & \numrange{0.3}{0.6}\,mm (\tip{hddg}) 2.5\,mm (\tip{ivdg})  & \numrange{0.1}{9.9}\,mm   & \numrange{0.2}{8.0}\,mm       \\ 
        Speed range      & \numrange{1}{8}\,m/s & all  & 0.2–20.0\,m/s    \\ 
        Diameter precision  & $\pm$0.03\,mm (\tip{hddg})        & $\pm$0.19\,mm    & $\pm$2\,mm for small  \\
        Speed precision  & $\pm$6\% (\tip{hddg})       & $\pm$4\%           & $\pm$5\%              \\ 
  
        \hline
    \end{tabular}
    \begin{tablenotes}
        \item[1]{DAVIS346 from \href{https://www.inivation.com}{www.inivation.com}, based on \textcite{Taverni2018-bsi-vs-fsi-davis} FSI sensor chip.}
            \item[2] \cite{kruger_two-dimensional_2002}
    \item[3]  \cite{ott2016operating}
    \end{tablenotes}
   \end{threeparttable}
\end{table}

\section{Conclusions} \label{sec:conclusions}
Our paper proposes an innovative way to measure droplets using an activity-driven \tip{dvs} event camera that observes the droplets falling through a shallow \tip{dof}. 
 Our results demonstrate the feasibility of this \tip{dvsd} method for droplets ranging from 0.3\,mm to 2.5\,mm, covering most of the real rainfall range.
 Droplet size and velocity measurements from the \tip{dvsd} have a maximum of 7\% \tip{mape} compared to the ground truth from the drop generator. 
 The droplet size uncertainties of the \tip{dvsd} measurements and \tip{gt} values are 10\% and 3\% respectively, 
 whereas the droplet velocity uncertainties are both 7\%. 
 The uncertainty of our prototype is encouraging because we expect substantial potential for improvement through more advanced hardware and processing methods. Most of all, our results are virtually unbiased, especially for small drops, which are difficult to observe for existing optical disdrometers.

With our strongest magnifying lens, our \tip{dvsd} prototype---under laboratory conditions---surpasses \tip{soa} disdrometers in terms of precision even though the sampling area is much smaller, as Table~\ref{tab:specs} shows. For future work, our aim is to increase the sensor resolution and capture real rainfall data with comparisons to \tip{soa} disdrometers.

  Today, the installation of multiple \tips{dvsd} would be expensive due to the prototype costs of the \tip{dvs} cameras. However, mass production for \tip{dvs} applications in consumer electronics will rapidly decrease production cost and improve the resolution and quality of \tip{dvs} cameras. 
  A \tip{dvsd} based on a low-power, inexpensive embedded Linux microcomputer could be developed that can autonomously estimate droplet diameters and velocities in real time while surviving harsh weather conditions in remote areas disconnected from the power grid. 
  The rain-driven computation and simple optical and lighting requirements of a \tip{dvsd} would be a great advantage compared to alternative optical disdrometers that sample at a constant high rate and require more complex optical and lighting arrangements. 


\section*{Data availability and Supplementary Material}
\label{sec:data_availability}
Our Supplementary Material details our materials and methods and our raw data and videos are available online\footnote{\href{https://drive.google.com/drive/folders/153C2YDQh-AFjdBd1kromg9BBv2esfq8e}{Raindrop measurements with an event camera - Public Google drive}}.


\section*{Author contributions}
\small{J. Steiner and K. Micev performed most of the experimental work and data analysis. A. Aydin performed initial experiments to establish the concept. T. Delbruck and J. Rickermann conceived and supervised the project. All authors participated in the writing of the paper.}

\section*{Competing interests}
The authors declare that they have no conflict of interest.

\section*{Acknowledgments}
\small{We thank G. Taverni for assistance with initial feasibility studies, S. Nasser, N. Ashgriz, and R. Loidl for their help with the \tip{hddg}.}







\AtNextBibliography{\small}
\printbibliography
\newpage
\setcounter{section}{0}
\setcounter{equation}{0}
\setcounter{figure}{0}
\setcounter{table}{0}
\setcounter{page}{1}
\setcounter{footnote}{0}
\makeatletter
\renewcommand{\thesection}{S}
\renewcommand{\thesubsection}{S.\arabic{subsection}}
\renewcommand{\theequation}{S\arabic{equation}}
\renewcommand{\thefigure}{S\arabic{figure}}
\renewcommand{\thetable}{S\arabic{table}}
\renewcommand{\thefootnote}{\textit{\alph{footnote}}}  
\glsresetall 
\newrefsection

\lhead{Raindrop measurement with event camera: Supporting information} 

\section{Supporting information: Methods and materials} \label{subs:sm}

The \MATLAB code used for data analysis, raw data, simulations, graphs, photos and videos of the experiments is available from the following Google drive link: 
\href{https://drive.google.com/drive/folders/153C2YDQh-AFjdBd1kromg9BBv2esfq8e}{Raindrop measurements with an event camera - Public}.

This supplementary material has the following sections:
\begin{itemize}
    \item \ref{subs:dvs_camera} provides details on the \tip{dvs} event camera.
    \item \ref{subs:drop_creation} describes our \tip{hddg} and \tip{ivdg}.
    \begin{itemize}
        \item \ref{subs:droplet_needle} details our droplet needle.
        \item \ref{subs:hddg_construction} details construction of the \tip{hddg}.
    \item \ref{subs:stability_of_hddg} provides additional useful methodology for testing the stability of the \tip{hddg}
        \item \ref{subs:hydrophobic_coatings} describes unsuccessful attempts to develop hydrophobic coatings to reduce the \tip{ivdg} droplet size.
    \end{itemize}
        \item \ref{subs:experimental_setup} details our experimental setups.
    \item \ref{subs:experiments} describes our experiments and the methods we used to measure droplets and estimate uncertainty, in particular
    \begin{itemize}
        \item \ref{subs:error_analysis} details our error analysis.
        \item \ref{subs:optimizing_lighting} explains how we optimized our droplet illumination.
        \item Secs. \ref{subs:calibration}, \ref{subs:angle}, and \ref{subs:height} explain our calibration of the optical magnification of the camera $M$, angle $\alpha$, and droplet fall height.
        \item \ref{subs:jaer_pics} explains how we measured the droplet diameter and speed in the image plane
        \item \ref{subs:jaer_measurement} provides formulas to compute the physical diameter and speed from the image plane measurements.
     \item \ref{subs:ground_truth} describes how we measure droplet flow rate for ground-truth droplet size estimates.
       \item \ref{subs:drop_calc} further details our \tip{gt} measurement of droplet mass.
    \end{itemize}
    \item \ref{subs:simulation} explains our model of the droplet speed versus fall height and droplet diameter, which we used to obtain our \tip{gt} speed from mass and to ensure that droplets fell at  close to terminal speed.
\end{itemize}

\subsection{Dynamic vision sensor event camera} \label{subs:dvs_camera}

\begin{figure}[!h]
    \centering
    \includegraphics[width=\textwidth]{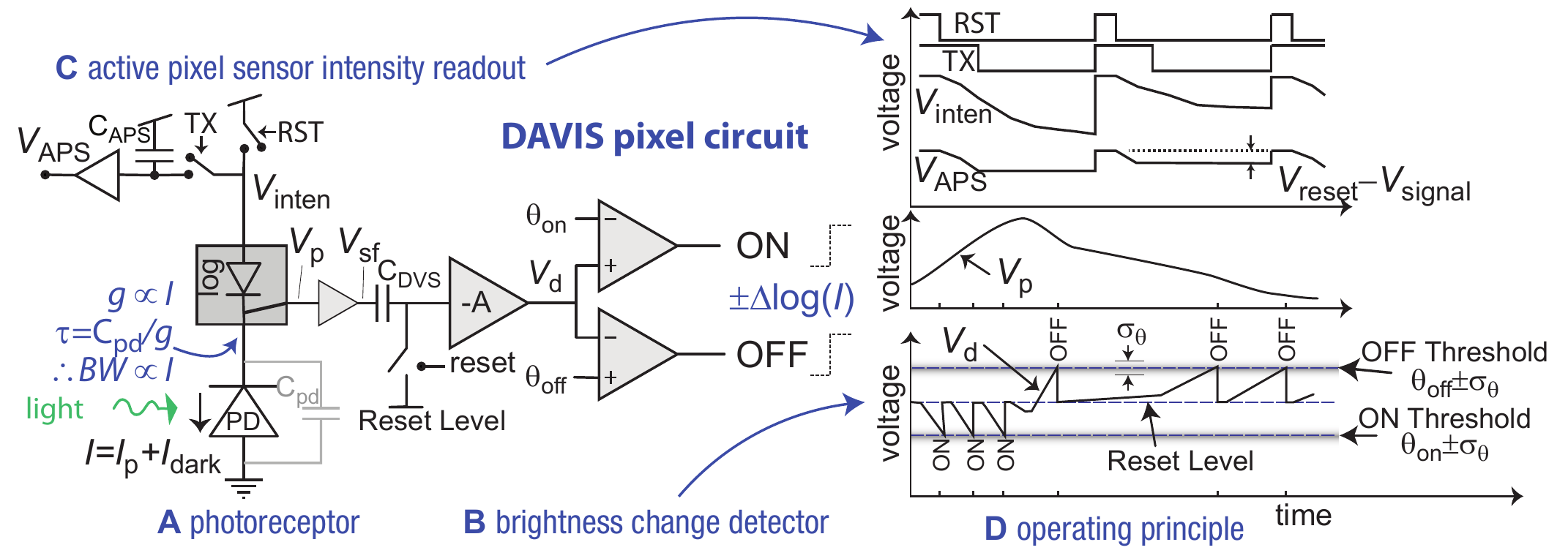}
    \caption{\tip{davis} pixel circuit and operating principle. The sensor generates asynchronous brightness change ON and OFF \tip{dvs} events and \tip{aps} intensity samples, which we do not use for disdrometry, but are helpful for calibration and focusing.}
    \label{fig:davis_pixel}
\end{figure}

The \tip{dvsd} uses a \tip{dvs} event camera.
Fig. \ref{fig:davis_pixel} shows the \tip{dvs} pixel circuit. Its design is based on the \tip{dvs} \parencite{Lichtsteiner2008-dvs} and the \tip{davis} \parencite{Brandli2014-davis} with improvements described in \textcite{Taverni2018-bsi-vs-fsi-davis}. 
It was developed by the Delbruck lab and is sold by inivation.com as the DAVIS346 camera. 
For the \tip{dvs} brightness change events used for drop measurement, the logarithmic photoreceptor (\textbf{A}) drives a change detector (\textbf{B}) that generates the ON and OFF events (\textbf{D}).
Pixel photoreceptors continuously transduce the photocurrent $I$ produced by the \tip{pd} to a logarithmic voltage $V_\text{p}$, resulting in a dynamic range of more than 120 dB. 
This logarithmic voltage (called \emph{brightness} here) is buffered by a unity-gain source follower to the voltage $V_\text{sf}$, which is stored in a capacitor $C_\text{DVS}$ inside individual pixels, where it is continuously compared to the new input. 
If the change  $V_\text{d}$ in log intensity exceeds a critical event threshold, an ON or OFF event is generated, representing an increase or decrease of brightness. The event thresholds $\theta_\text{on}$ and $\theta_\text{off}$ are nominally identical for the entire array.
The time interval between individual events is inversely proportional to the derivative of the brightness. When an event is generated, 
the pixel’s location and the sign of the brightness change are immediately transmitted to an arbiter circuit surrounding the pixel array, then off-chip as a pixel address, and a timestamp is assigned to individual events. The arbiter circuit then resets the pixel’s change detector so that the pixel can generate a new event. 
Events can be read out at up to rates of about 10\,MHz. The quiescent (noise) event rate is a few kHz. Events are transmitted from the \tip{davis} chip to a host computer over \tip{usb}. 
The host software records the data and allows playback in slow motion. In addition to the \tip{dvs} circuit the DAVIS346
also has a circuit for conventional intensity frame recordings called the \tip{aps} circuit (Fig.~\ref{fig:davis_pixel}C),
which was useful
for lens calibration and focusing.

\subsection{Droplet generation} \label{subs:drop_creation}

Figs. \ref{fig:fig1}D and \ref{fig:hddg_sketch} illustrate our \tip{hddg} and Fig.~\ref{fig:experimental_setup_HDD} shows photos of the \tip{hddg} setup. 
Our \tip{hddg} is based on previous work by \textcite{Kosch2015-hdd-droplet-generator}, who utilized a computer hard disk arm as an actuator.
They used a high-frequency buzz to create ripples in a steady stream of water emitted by a stiff glass needle, which would break up into small droplets. Our \tip{hddg} uses a soft and flexible plastic needle, 
which, if properly combined with a controlled stream of water, 
reliably creates a single droplet at each end of an oscillation, 
resulting in two droplet streams, one of which we measured. 

\subsubsection{\tip{hddg} droplet needle} \label{subs:droplet_needle}
To make the droplet needle, we used a microloader microcapillary tip\footnote{OD 0.3\,mm Eppendorf 20\,ul microcapillary pipette; \href{https://online-shop.eppendorf.us/US-en/Manual-Liquid-Handling-44563/Pipette-Tips-44569/Microloader-PF-56180.html\#isProductInfo}{Merck Catalog No. 930001007}}
This soft plastic needle tubing protrudes from its integrated feeder expansion.
The peristalitc pump tubing\footnote{OD 3.5\,mm Tygon® S3™E-3603, Saint-Gobain Performance Plastics; \href{https://www.tubes-international.com/products/industrial-hoses-delivery-and-suction-hoses/tygon-hoses-and-tubings/}{Tygon tubing website}} 
is plugged into this microloader. 
The needle is threaded through a hole drilled through the \tip{hdd} actuator arm so that the needle protrudes from the arm by \SIrange{2}{4}{cm}, and we can control the length of the protruding needle to adjust its resonance frequency to match the driving frequency. That way, we can use a smaller driving voltage and current for the HDD driver coils. The needle is fixed to an elevated platform with a conical interference fit between the needle and a black plastic tube glued to that platform (see Fig.~\ref{fig:experimental_setup_HDD} C and D).
The \tip{hdd} arm is actuated with an audio power amplifier driven by sinusoidal waveforms generated by an audio wave generating program (\href{https://www.szynalski.com/tone-generator/}{www.szynalski.com/tone-generator}). The arm is coupled to the needle by threading the needle through one-sided adhesive tape which is applied over the hole in the arm.
We used amplitudes from
\SIrange{3}{10}{Vpp} 
and frequencies from 
\SIrange{60}{220}{Hz}. By adjusting the flow rate and oscillation frequency,
we can arrive at a combination of settings where nearly on every oscillation, 
a single droplet is flung from the needle tip at each end of the oscillation.
Since the flow rate and frequency are constant, the droplet sizes are also constant.

To create large droplets, we used an \tip{ivdg} assembled from a standard IV dripper
and a needle tip intended for glue dispensers\footnote{ID 0.2\,mm, OD 0.4\,mm, part VD90.0032 \url{https://www.martin-smt.de}}.
The diameter of the needle has only a weak influence on the droplet size, 
which is mainly determined by the surface tension of water adhering to the needle; at low flow rates, when the droplet mass grows large enough, it breaks free 
from the needle.
We adjusted the IV flow to produce a regular series of droplets.

\subsubsection{Construction of \tip{hddg} droplet generator}\label{subs:hddg_construction}
\begin{figure}[!h]
    \centering
    \includegraphics[width=\textwidth]{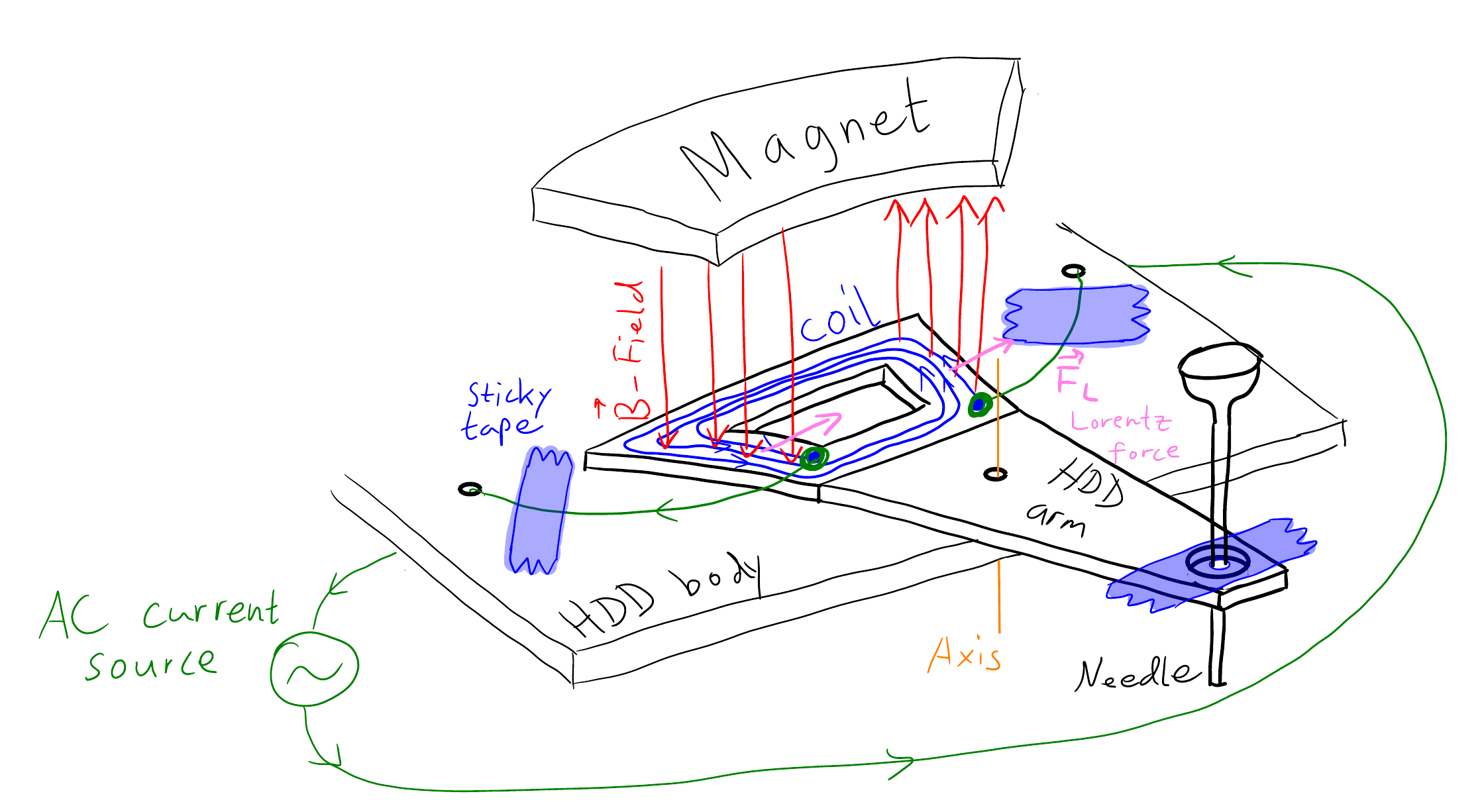}
    \caption{\textbf{Sketch of \tip{hddg}.} The red B-vector is the induced magnetic field by the coil and alternates its direction from up to down. When it points up, it is attracted to upwards pointing yellow B-vectors from the static magnet. Two sticky tapes and two green wires hold the \tip{hdd} arm in place in a spring-like manner. This attachment ensures that the coil is centered between the two opposite poles of the metal piece. Without the sticky tapes, we observed that the arm drifts to one side and does not function properly anymore.}
    \label{fig:hddg_sketch}
\end{figure}

Fig.~\ref{fig:hddg_sketch} sketches the \tip{hddg} construction. We used an old 250GB 3.5" hard disk drive that we disassembled to expose the platter head actuator arm.
The copper wires have two functions: first, to power the \tip{hdd} coils and second, to act like springs to keep the \tip{hdd} arm close to the middle of the two magnets, which is the best operating point for the arm. To construct the needle driver, we follow these steps:
\begin{enumerate}
\item The upper end of the plastic needle was frictionally held in place (see black tube).
\item Two nuts and a bolt (Fig.~\ref{fig:experimental_setup_HDD}D) could adjust the protruding length of the needle to match the resonance frequency of the needle with the \tip{hdd} frequency to maximize the amplitude of the oscillation and, therefore, the efficiency.
\item The lower end of the plastic needle was guided through a tiny hole in a piece of sticky tape. The tape was attached to the end of the \tip{hdd} arm.
\item The needle was connected to the water tube from the pump. This connection must be tight to prevent water leakage and to prevent the needle from twisting.
\item Since the needle has some inherent curvature, we twisted it with our fingers until the inherent curvature was perpendicular to the oscillation direction.
\end{enumerate}

\subsubsection{Testing the stability of the hard disk drop generator with a strobe light} 
\label{subs:stability_of_hddg}

It can be observed in our sample recording video that the \tip{hddg} droplet jet always changed direction a bit. It seemed that the jet moved in "waves" and "cycles". So we had to be patient and lucky that the jet landed in the DVS measurement area. This is what we meant by ``unstable drop generation by \tip{hddg}'' in Sec.~\ref{sec:discussions}.  There are probably several effects that caused this behavior, \ie  irregular water flow from the pump, loosely attached needle (so that it can rotate and tilt slightly) and air currents. We believe that air currents only played a minor role and that the most likely culprit is the circular hole in the \tip{hdd} arm that couples it with the needle (Fig.~\ref{fig:experimental_setup_HDD}E).

Under our lighting and with \tip{hddg} drop generation frequencies above 20\,Hz, any errors with the \tip{hddg} drop release are too fast to be seen with the naked eye. Thus, to test the \tip{hddg} drop release, we used a custom-built strobe light to illuminate the oscillating needle. For the initial experiments, the goal was to create one drop per cycle, which meant that the drop stream was one-sided. For higher frequencies, it was impossible to tell if a drop is released every one or every two cycles. We built an Arduino-driven LED strobe light to check the drop creation frequency using the principle of aliasing. We set the strobe light frequency to the same as the desired drop creation frequency for two seconds, and for another two seconds, the strobe frequency was half of the desired drop creation frequency. If the distance between successive droplets (see Fig.~\ref{fig:experimental_setup_HDD}F) changes between the droplets when illuminated with the two different strobing frequencies (i.e. distance doubles when the strobe light is flashing at half the desired drop creation frequency), only one drop is released every second cycle. If the distance between the droplets stays the same, one droplet is released per cycle, which is desired.

\subsubsection{Testing different hydrophobic coatings for the \tip{ivdg}} 
\label{subs:hydrophobic_coatings}

Using a basic \tip{ivdg} generates different drop sizes that are created depending on the needle size, water surface tension and needle materials. The bigger the needle orifice, the larger the drops. However, this is limited at some point since small drops stick to the needle due to attraction between water and needle tip. The needle surface can be made water repellent by applying a hydrophobic coating. This is rather difficult to do, since the needle orifice is often too small for this coating to fully enter, and the coating is also washed away after a few minutes by the water itself. Also, we did some tests with soap to lower the surface tension of the water. The results did not look promising, which prompted us to develop a droplet generator that can reliably generate drops of the desired size. 

\subsection{Experimental setups} \label{subs:experimental_setup}

\begin{figure}[!h]
    \centering
    \includegraphics[width=\textwidth]{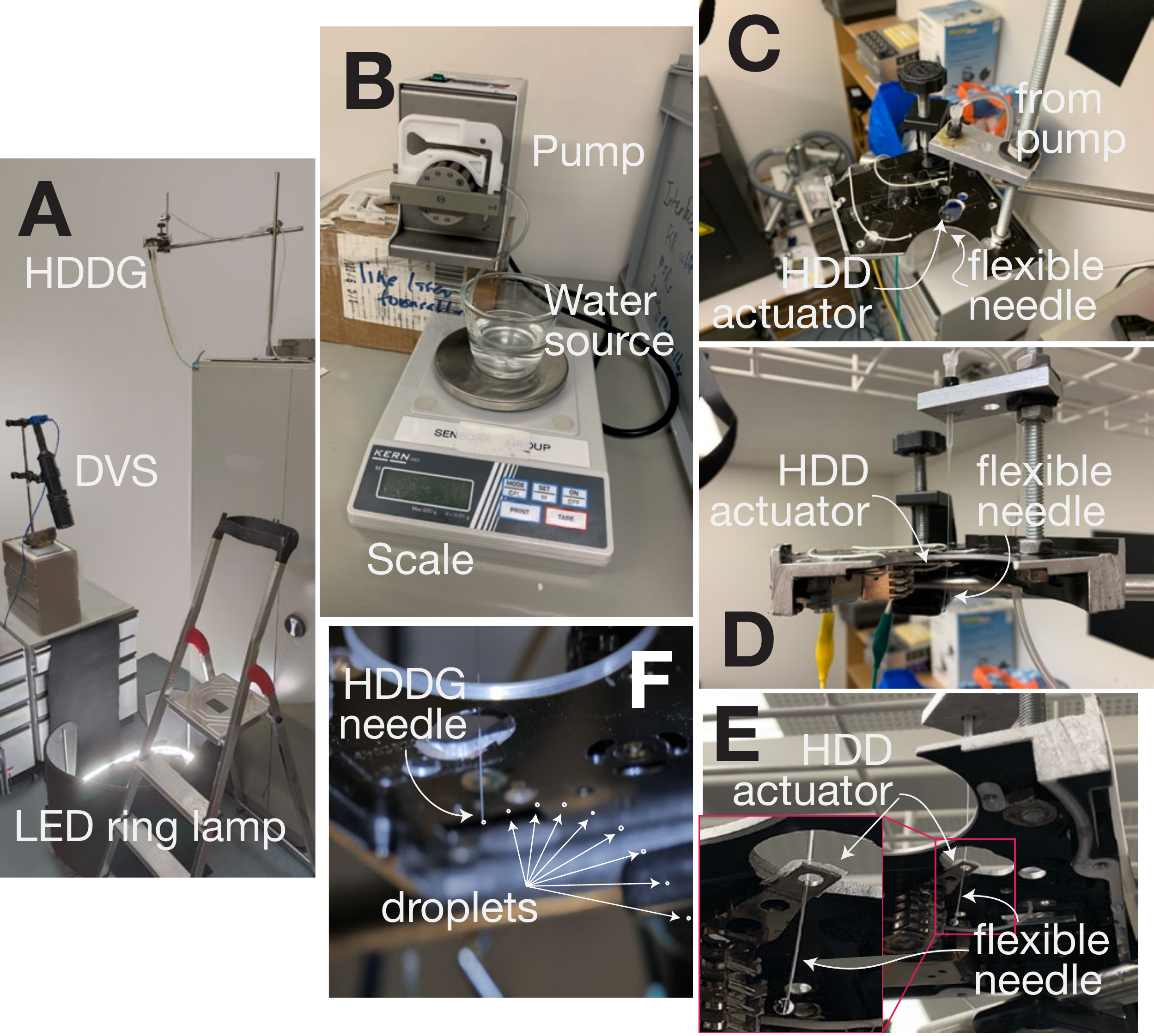}
    \caption{Pictures of the \tip{hddg} experimental setup. \textbf{A}: Overview of the whole \tip{hddg} setup. \textbf{B}: Peristaltic pump, scale and water tank. \textbf{C, D, E}: \tip{hddg} drop generator from three different perspectives. \textbf{F}: Stream of droplets created by the \tip{hddg}.}
    \label{fig:experimental_setup_HDD}
\end{figure}

\begin{figure}[!h]
    \centering
    \includegraphics[width=\textwidth]{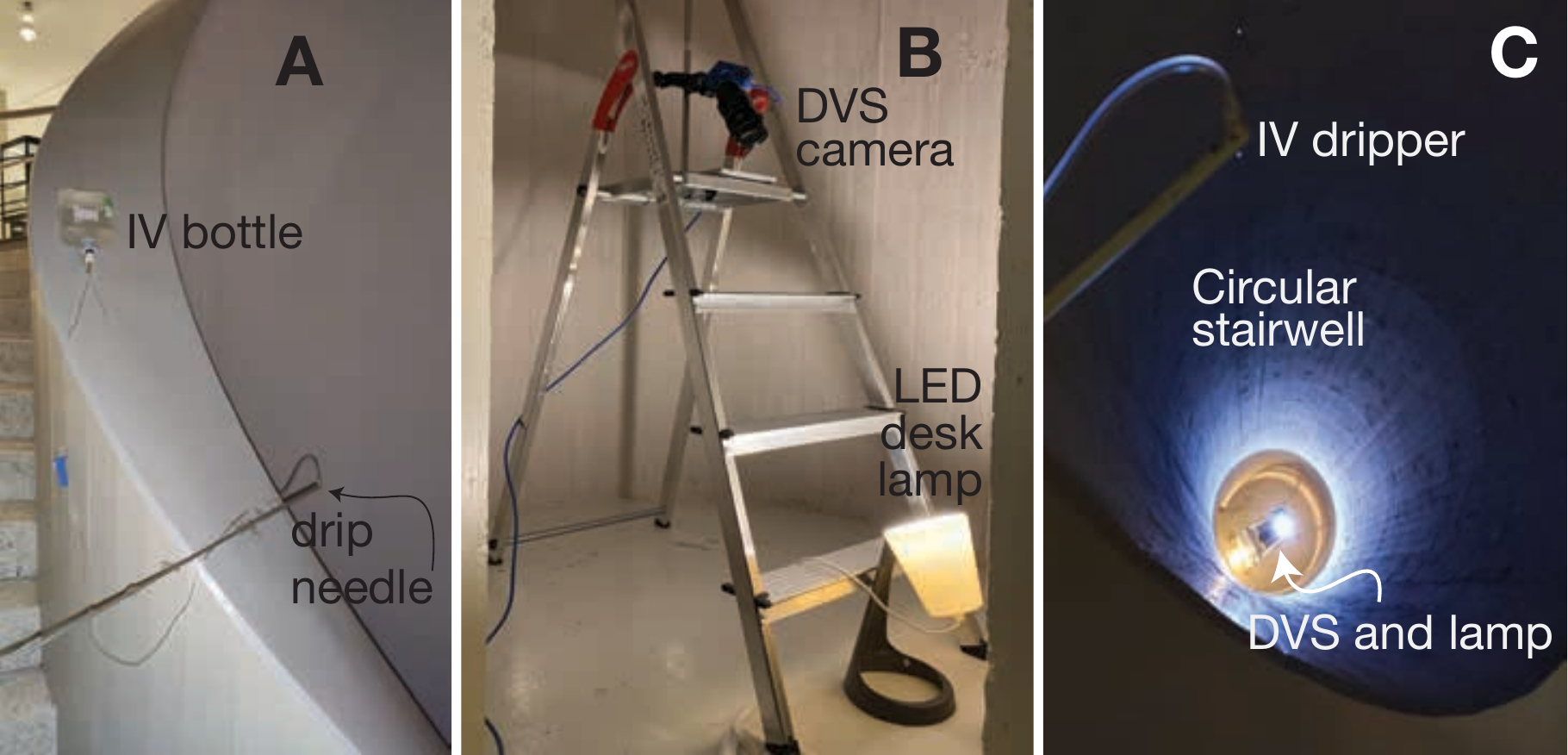}
    \caption{Pictures of the \tip{ivdg} experimental setup. \textbf{A}: \tip{ivdg} drop generator with water source and rod holding the needle. \textbf{B}: \tip{dvs} camera and lighting setup. \textbf{C}: Tube inside the staircase with a fall height of 10\,m with the \tip{dvs} camera and lighting setup at the bottom.}
    \label{fig:experimental_setup_IV}
\end{figure}

The \tip{hddg} experiment is shown in Fig. \ref{fig:experimental_setup_HDD}, while images of the \tip{ivdg} experiment can be found in Fig. \ref{fig:experimental_setup_IV}.

\begin{figure}[h!]
    \centering
    \includegraphics[scale=0.03]{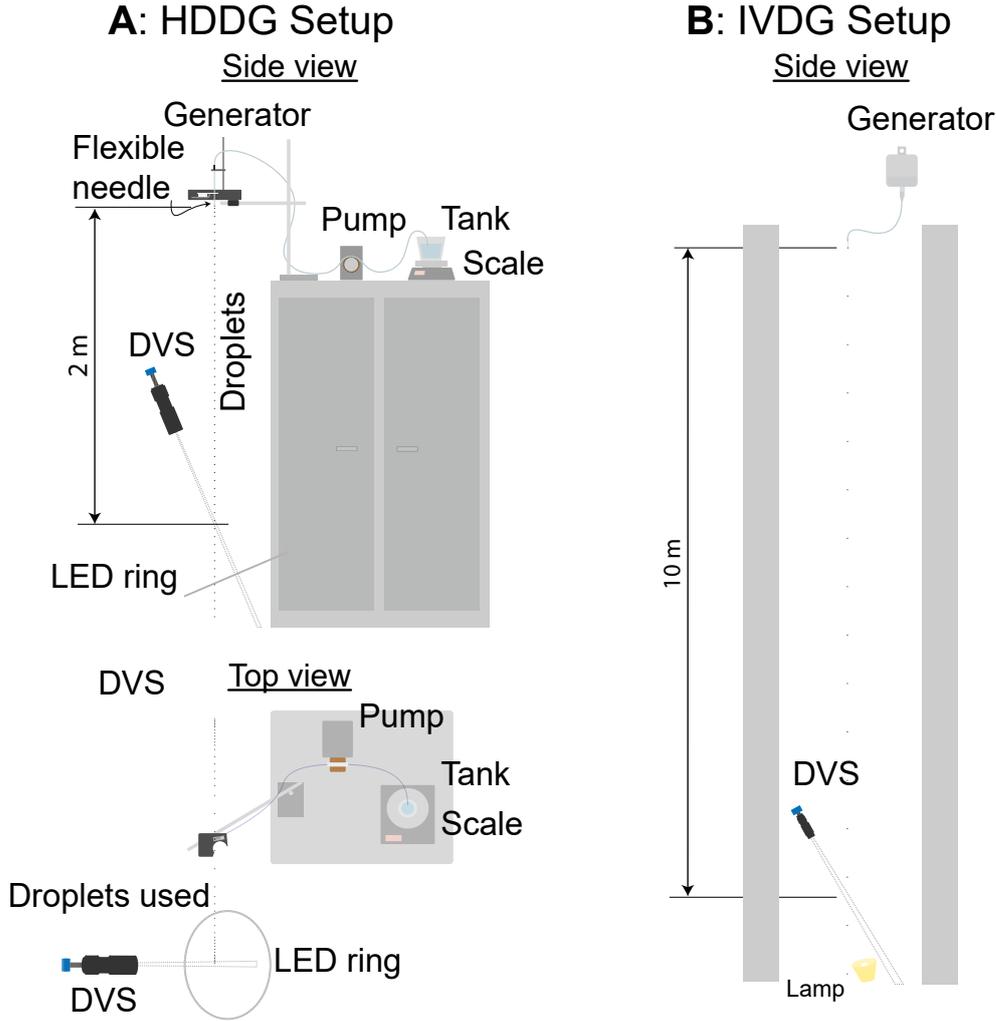}
    \caption{\textbf{Illustration of experimental setups.} \textbf{A:} \tip{hddg} setup is illustrated from a side view and a from a top view perspective. \textbf{B:} \tip{ivdg} setup is illustrated from a side view perspective.}
    \label{fig:setup}
\end{figure}

Fig. \ref{fig:setup} illustrates the \tip{hddg} setup (left side) and \tip{ivdg} setup, from a side view perspective. The \tip{hddg} setup is additionally shown from a top view in Fig. \ref{fig:setup} (bottom left corner) .

\subsection{Experiments} \label{subs:experiments}

The objective of the experiments was to measure the diameter and velocity of droplets with a \tip{dvs} with droplets falling close to their terminal speed, and to compare these measurements with their corresponding \tip{gt}. Two experiments were conducted with different setups to optimize droplet creation for two different drop size categories, which are further explained in the following two paragraphs. Both experiments used the exact same \tip{dvs} camera and measurement principles (described in \ref{subs:jaer_measurement}). Both lenses were set at their minimum focusing distance. The working distance was then reduced further to about 50\,cm using lens spacers. The \tip{dvs} was a \tip{davis} (model: DAVIS346), designed by the Delbruck group and manufactured by iniVation. It has a resolution of $346\times260$ pixels and a pixel array size of $8\times6$\,mm$^2$ \autocite{Brandli2014-davis,Taverni2018-bsi-vs-fsi-davis}. Pixels have a pitch of 18.5\,$\mu$m. Table \ref{tab:experiment_setup} compares both experimental setups with further details. Both droplet generation methods are described in \ref{subs:drop_creation}.

The first experiments, also called \tip{hddg} experiments, were conducted in a darkroom using an \tip{hddg} and a fall height of 2\,m. 
This setup was optimized for droplets with a diameter of 0.3 to 0.6\,mm, by using a 300\,mm lens with the resulting large magnification $M=30.7\,\text{px/mm}$. This led to a sampling area of $11.2\times 8.4$\,mm$^2$. 
The \tip{hddg} produced a localized drop jet, 
making it fairly easy to hit the sampling area. The height of the droplet fall is enough for the drops to reach more than 99\% of the terminal velocity according to \ref{subs:simulation}. 
We used a 40\,W \tip{led} ring purchased from a home supply store as a light source pointing upward and inward to achieve a high contrast between the bright drop edges and the dark background (see Fig. \ref{fig:experimental_setup_HDD}A).

The second experiment, also called the \tip{ivdg} experiment, was carried out in the vertical tunnel of a spiral staircase using a \tip{ivdg} as a droplet generator and a fall height of more than 10\,m. With a smaller magnification of 7\,px/mm (4.4$\times$ smaller than for the \tip{hddg} experiments), the magnification was reduced for droplets with a diameter of approximately 2.5\, mm, while maintaining a relative precision similar to that of the \tip{hddg} setup. The main reason for this reduction in magnification was to more easily capture the droplets, which had a huge scatter compared to the \tip{hddg} scatter. With the \tip{ivdg}, it was only possible to create a single droplet size because the droplet size is determined by the weight that breaks the surface tension with the needle. The higher magnification led to a larger sampling area of roughly $49\times37$\,mm$^2$, which increases the chances to capture the larger drops, which have a much greater spread from the higher fall height needed to achieve a final speed close to the terminal speed. The height of the fall is sufficient for the drops to reach more than 97\% of the terminal velocity according to \ref{subs:simulation}. The working space of the experiment \tip{ivdg} was more limited (see Fig. \ref{fig:experimental_setup_IV}B), so we used a single 5\,W \tip{led} reading lamp to illuminate the drops from behind, which refracted the light towards the middle of the drop, leaving the edges dark. A large contrast was achieved between the bright background and the dark edges of the drop.

A key factor was the adjustable angle of the camera $\alpha$ (see Fig. \ref{fig:fig1}B: left). The smaller is $\alpha$, the more accurately the diameter can be measured and the larger is the sampling area, which leads to a faster estimate of \tip{dsd}. Larger $\alpha$'s allow more precise velocity measurements, but smaller sampling area. The sampling area is the total area of the \tip{pof} inside the \tip{fov} multiplied by $\cos{\alpha}$. According to our findings, 20° to 30° was optimal for $\alpha$. 

Eq. \eqref{eq:d_DVS} was used for the diameter calculation. For the velocity calculation the non-simplified formula on the left side of \eqref{eq:v_DVS} was used for the \tip{hddg} experiment, while the simplified formula on the right side of \eqref{eq:v_DVS} was used for the \tip{ivdg} experiment due to the negligible horizontal velocity component on the recording (see Fig. \ref{fig:fig1}B: right). 

Another important requirement was a flicker-free DC light source, which would otherwise create flicker artifacts in the recording. Pictures and illustrations of the experimental setups are provided in \ref{subs:experimental_setup}.

\subsubsection{Data selection}\label{subs:data_selection}
Each recording session collected data for a single droplet size. From the recording, we manually selected droplets that passed from the top to the bottom of the image and created a distinct hourglass shape. Fig.~\ref{fig:fig1_methods} illustrated how this procedure excluded droplets that did not pass through the \tip{pof}.

\begin{table}[h!]
    \centering
    \begin{tabular}{l|cc}
        \multirow{2}{*}{Setup parameters}        & \multicolumn{2}{c|}{Experiment}      \\
                                & \tip{hddg}                      & \tip{ivdg}                      \\
        \hline
        Drop diameter           & \numrange{0.3}{0.6}\,mm   & 2.5\,mm                   \\
        Drop velocity           & \numrange{1.4}{2.2}\,m/s  & 7.7\,m/s                  \\
        Fall height             & 2\,m                      & 10\,m                     \\
        Lighting          & 40\,W LED ring                  & 5\,W LED bulb                  \\
        Total luminance         & 4500\,lm                & 470\,lm                   \\
        Location                & darkroom                  & spiral staircase          \\
        Lens                    & TAIR-3 (Russian sniper rifle)                   & Edmund manual zoom        \\
        Listed focal length     & 300\,mm                   & 75\,mm                    \\
        Aperture ratio          & f/4.5                     & f/1.2                     \\
        Lens Spacer                 & M42-C (16\,mm long)       & -                         \\
        \# 5\,mm C-CS lens spacers   & 23                        & 2                         \\
        Working distance        & 50\,cm                    & 50\,cm                    \\
        Sampling area           & 11$\times$8\,mm$^2$       & 49$\times$32\,mm$^2$      \\
        Camera \tip{fov} angle $\alpha$   & 22°                       & 29.5°                     \\
        Magnification $M$       & 30.7\,px/mm               & 7.0\,px/mm                \\

        \hline
    \end{tabular}
    \caption{Detailed setup parameters for the \tip{hddg} and \tip{ivdg} experiment.}
    \label{tab:experiment_setup}
\end{table}

\subsubsection{Error analysis} \label{subs:error_analysis}

In both experiments, two aspects were considered to determine the propagation of the error of the measurements. 

The first aspect is the combined measurement uncertainty of the \tip{dvsd} and \tip{gt} values which factors in all uncertainties. For example, one measurement uncertainty of the \tip{dvsd} comes from the limited sensor resolution of the \tip{dvsd}. Another one comes from the uncertainty of $\alpha$. Together with other uncertainties, we could then calculate the combined standard uncertainty of the DVSD velocity and diameter measurements. The combined \tip{gt} uncertainty consists of the uncertainty in height, droplet mass, drop creation frequency, etc. According to \textcite{jcgm_guide_2008}, in the case of uncorrelated input quantities $x_i$, the combined standard uncertainty of a function $u(f)$ can be described as:
\begin{equation}
    u(f) = \sqrt{\sum_{i=1}^{N}{ \left[\pdv{f}{x_i} u(x_i)\right]^2 }} = \sqrt{ \left[\pdv{f}{x_1}u(x_1)\right]^2 + \left[\pdv{f}{x_2}u(x_2)\right]^2 + \left[\pdv{f}{x_3}u(x_3)\right]^2 + \ldots}
\end{equation}
where $u(x_i)$ is the standard uncertainty of the input quantity $x_i$. The uncertainty is either derived with a \textit{Type A evaluation}, where the standard uncertainty is evaluated with the experimental standard deviation from repeated observations, or with a \textit{Type B evaluation}, where the estimated uncertainty is evaluated using our judgement of  uncertainty. 
We used a \textit{Type B evaluation}, either by using the manufacturer's instrument specifications or by a conservative estimate of the measured uncertainty. A linear approximation of the function is used for each input quantity.
A spreadsheet in our online results folder\footnote{\href{https://drive.google.com/drive/folders/15POK3C3b3o7tKXAPDlq94tO2Y-V1VmXx}{Results computations; see \textsf{00 README.txt}}} reports the final uncertainty values computed by our Matlab scripts.

The second aspect is accuracy, \ie, comparison of the measured values from the \tip{dvs} with the \tip{gt} values. This is done by calculating the \tip{mape}:
\begin{equation}
    \text{MAPE} = \frac{100\%}{n} \sum_{i=1}^{n} \left| \frac{GT_{i}-ME_{i}}{GT_{i}} \right|
\end{equation}
where $GT_i$ is the \tip{gt} value and $ME_i$ is the measured value.

\subsubsection{Optimizing the lighting}
\label{subs:optimizing_lighting}

\begin{figure}[h!]
    \centering
    \includegraphics[scale=1]{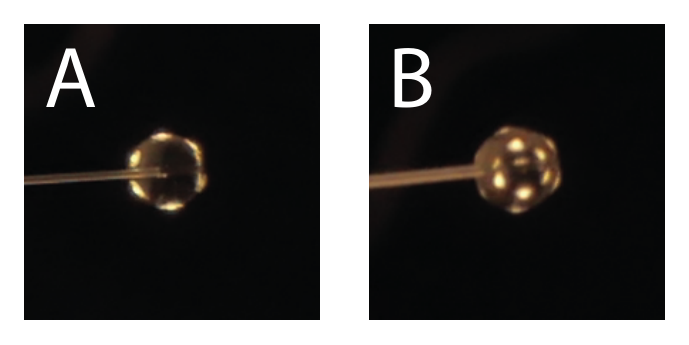}
    \caption{Drop illumination with \textbf{A} showing a drop with pronounced edges and \textbf{B} showing a drop with less pronounced edges.}
    \label{fig:refraction}
\end{figure}

Water droplets refract light in the same way that convex glass elements do, namely towards the middle. This property of drops was used to determine a good lighting setup for \tip{dvs}. To test this optical phenomenon, we took a dispensing needle and attached a tiny water drop at its end and took a photograph of it. We altered the distance between the circular light source until we were satisfied with the brightness of the drop edge. Fig. \ref{fig:refraction} shows the drop illumination for two different distances from the light source. In Fig. \ref{fig:refraction}A the edges of the drop are clearly pronounced, whereas in Fig. \ref{fig:refraction}A the edges are are very weak. It is therefore important to align the light source correctly, so that the edges of the droplet are well pronounced at the \tip{pof}. The light source used for this example was a ring light configuration with 6 \tip{led} bulbs. The light source used for the \tip{hddg} experiments was a ring light with a \tip{led} strip was used. However, the phenomenon is still the same. For the \tip{ivdg} setup we used a single LED desk lamp.

\subsubsection{Measurement of camera angle $\alpha$} \label{subs:angle}

The camera angle $\alpha$ was measured by aligning an iPhone Xs flush with the back of the \tip{dvs} body and reading off its 3D-orientation from the \tip{imu} accelerometer (using the Apple bubble level app). Before the measurement, we verified that the angle read zero when the phone was held on a flat surface.

\subsubsection{Measurement of fall height} \label{subs:height}

The fall height between the \tip{ivdg} and the \tip{dvs} sampling area was measured using a long wire, and the fall height between the \tip{hddg} and the \tip{dvs} sampling area was measured using a 2\,m folding ruler.

\subsubsection{Image plane droplet diameter and velocity measurement from \tip{dvs} recording} \label{subs:jaer_pics}

The droplet diameter and velocity are measured from the \tip{dvs} recording with jAER (\url{https://jaerproject.net}). The \textit{Speedometer}
\footnote{
\href{https://github.com/SensorsINI/jaer/blob/master/src/ch/unizh/ini/jaer/projects/minliu/Speedometer.java}{Speedometer class}.
\href{https://docs.google.com/document/d/1fb7VA8tdoxuYqZfrPfT46_wiT1isQZwTHgX8O22dJ0Q/edit\#bookmark=id.vwjmec8vrlvh}{Speedometer usage}.
}
plugin filter allows for convenient measurement of the diameter and velocity by outputting the velocity seen on the recording $v_\text{r}$ [kpx/s], horizontal displacement $\Delta x_\text{r}$ [px] and vertical displacement $\Delta y_\text{r}$ [px].

An hourglass appears on a \tip{dvs} recording after accumulating all events from one droplet passing through the \tip{fov}. 
The width at the center of the hourglass indicates the diameter of that water drop when in focus
(see Fig. \ref{fig:fig1}C on the right). 
This width $d_\text{r} \text{ [px]}$ is measured with
\textit{Speedometer} 
using two mouse clicks. 
A sample recording of a \tip{hddg} drop with a drop creation frequency $f$ of 100\,Hz (2$\times$50\,Hz, two-sided) is analyzed. Two drops are visible, but only the second drop on the right side is analyzed.

\begin{SCfigure}
    \centering
    \includegraphics[width=.5\textwidth]{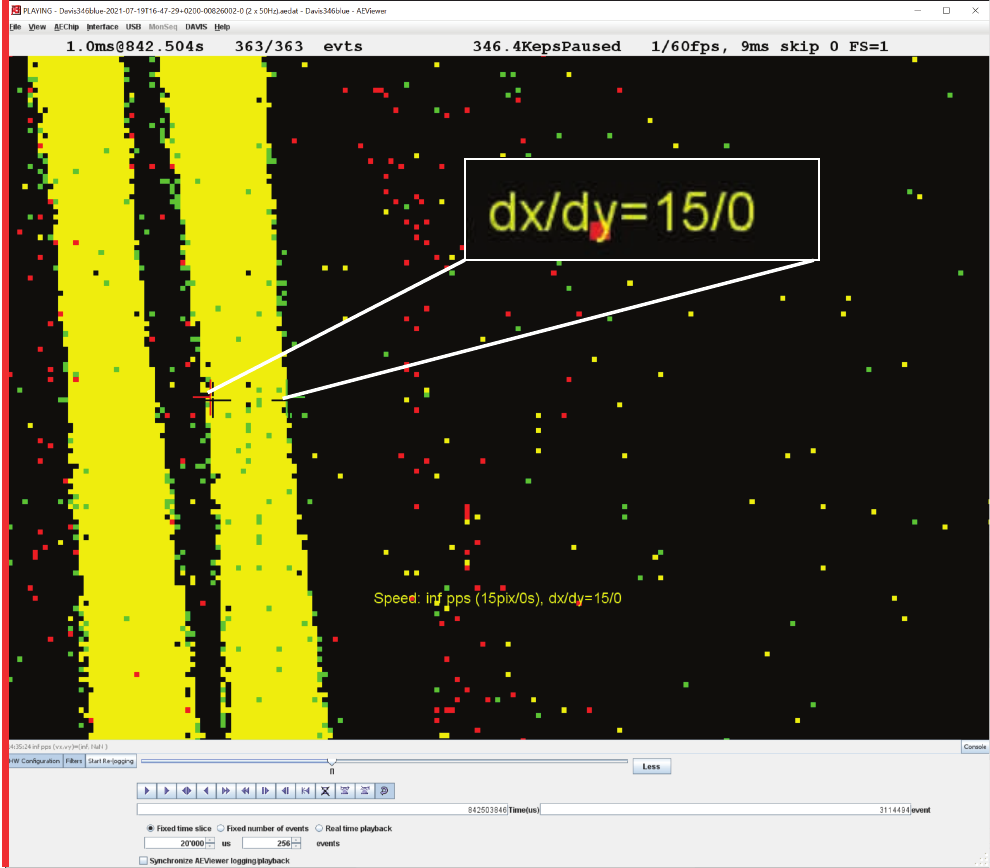}
    \caption{Measurement of the diameter from the \tip{dvs} recording using jAER. Two droplets passed during the accumulation time. After marking the left and right sides of the hourglass waist, \textit{Speedometer} displays the diameter $d_\text{r}$ [px]. This is a sample recording of an \tip{hddg} drop with a drop creation frequency $f$ of 100\,Hz (2$\times$50\,Hz, two-sided).}
    \label{fig:jaer_diameter}
\end{SCfigure}

Fig. \ref{fig:jaer_diameter} shows the diameter measurement. The right point at the thinnest width of the \textit{hourglass} is selected first (see Fig. \ref{fig:jaer_diameter}A). Next the left point on the thinnest width of \textit{hourglass} is selected (see Fig. \ref{fig:jaer_diameter}B).  \textit{Speedometer} displays the diameter of the recording $d_\text{r} = 15$\,px. The diameter $d$ [mm] can then be calculated with \eqref{eq:d_DVS}.

\begin{figure}
    \centering
    \includegraphics[width=\textwidth]{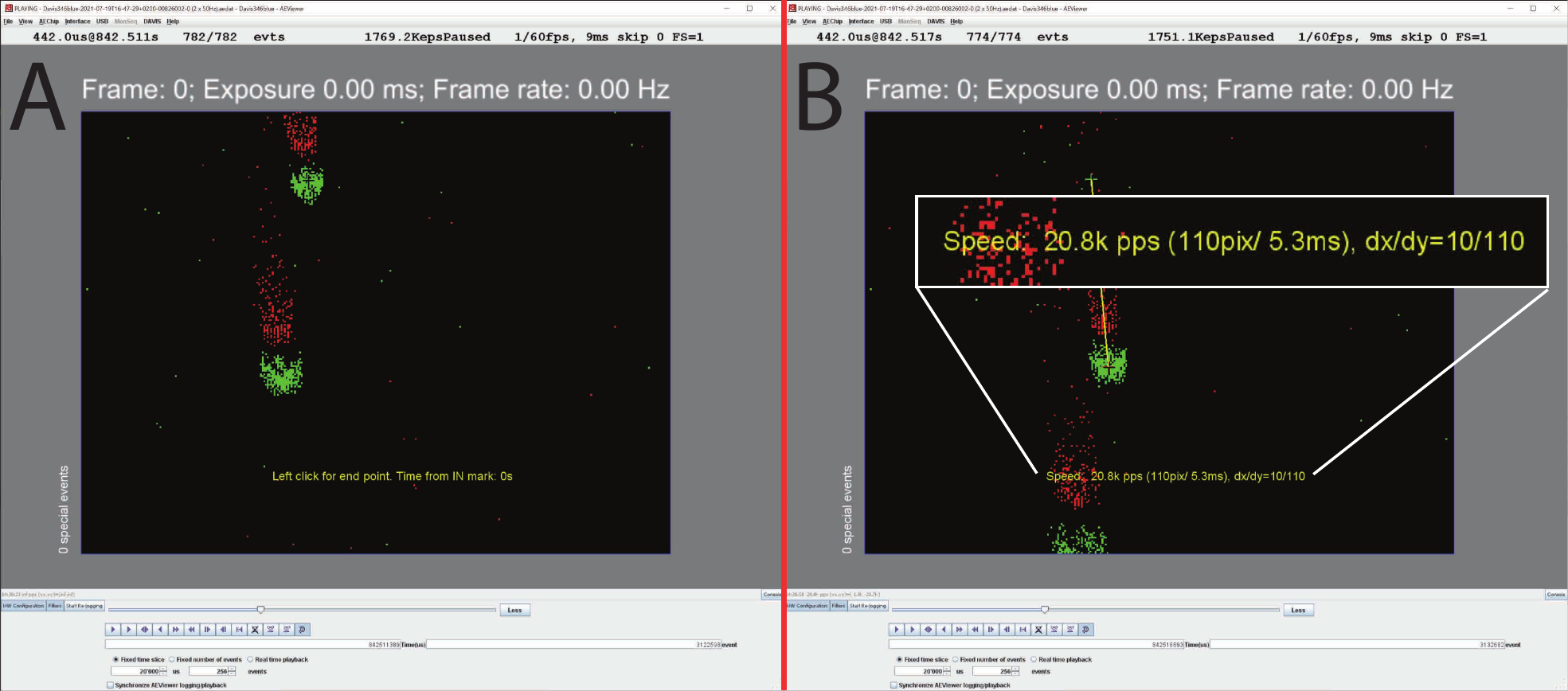}
    \caption{Measurement of the velocity from the \tip{dvs} recording using jAER. \textbf{A}: Mark IN point first on the mid-point of the right drop before the drop passes the \tip{pof}. \textbf{B}: Mark OUT point on the mid-point of the drop after the right drop passes the \tip{pof} and \textit{Speedometer} outputs the velocity $v_\text{r}$ [kpx/s]. This is a sample recording of an \tip{hddg} drop with a drop creation frequency $f$ of 100\,Hz (2$\times$50\,Hz, two-sided).}
    \label{fig:jaer_velocity}
\end{figure}

Fig. \ref{fig:jaer_velocity} shows the velocity measurement. Before the drop reaches the \tip{pof} (smallest diameter), the midpoint of the drop is selected (see Fig. \ref{fig:jaer_velocity}A). After that, the midpoint of the drop is selected again after the drop has passed the \tip{pof} (see Fig. \ref{fig:jaer_velocity}B). This is done by scrubbing through the recording. The \textit{Speedometer} outputs the velocity of the droplet in the image $v_\text{r} = 20.8$\,kpx/s. The physical velocity $v$ can then be calculated directly with Eq. \eqref{eq:v_DVS} (right side). For a more accurate calculation, the time difference $\Delta{t}=5304$\,µs can be used together with the horizontal and vertical displacement $\Delta x_\text{r}=10$\,px and $\Delta y_\text{r}=110$\,px to calculate the velocity $v$ according to  \eqref{eq:v_DVS} (left side). 

Calculations for diameter $d$ [mm] and velocity $v$ [m/s] are described in \ref{subs:jaer_measurement}.

\subsubsection{Optical and geometrical calibration}
\label{subs:calibration}
Before the diameter and velocity measurement with the \tip{dvs} can be performed,
the camera geometry and optics must first be calibrated. First, the \textit{camera angle} $\alpha$, which is the angle between the \tip{los} of the camera and the vertical $yz$-plane (see Fig. \ref{fig:fig1}B: left), is measured using a smartphone accelerometer (see \ref{subs:angle}). 
The \textit{droplet angle} $\beta$ is the angle between the projected droplet trajectory (trajectory in camera image) and the camera image $y_\text{r}$-axis on the \tip{dvs} recording (see Fig. \ref{fig:fig1}B: right). 
The magnification $M$ describes how large a certain distance in reality on the \tip{pof} would appear on the \tip{dvs} recording and vice versa. 
The magnification calibration is done by recording a miniature checkerboard calibration chart held at the \tip{pof} with the \tip{dvs}. 
The magnification $M$ is measured by dividing the checkerboard square size in mm by the number of pixels.

\subsubsection{Field of view}
\label{subs:fov}

\begin{figure}[h!]
    \centering
    \includegraphics{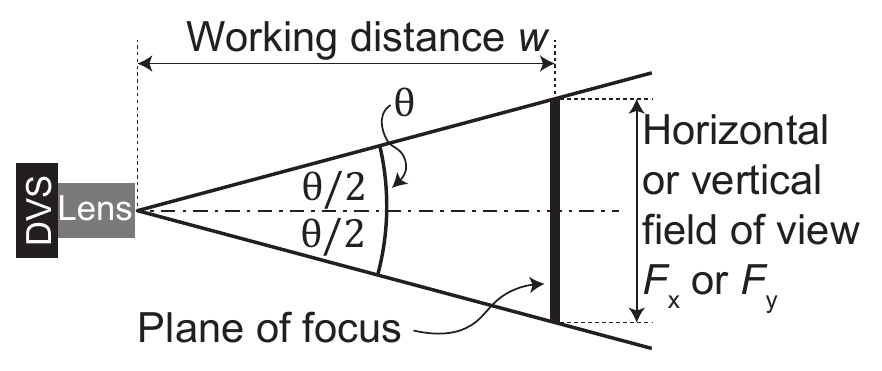}
    \caption{Illustration of the \tip{fov} $F_\text{x}$ and $F_\text{y}$ of the camera including the corresponding \tip{aov} $\theta_{\text{x}}$ and $\theta_{\text{y}}$.}
    \label{fig:fov}
\end{figure}

The geometry of the \tip{fov} and \tip{aov} is important for measuring the velocity on the \tip{dvs} camera. The illustration is can be seen in Fig. \ref{fig:fov}. The \tip{aov} $\theta$ is an important quantity, which can be calculated with the working distance $w$, horizontal \tip{fov} $F_\text{x}$ and vertical \tip{fov} $F_\text{y}$, that are both defined at the \tip{pof}. This is done as follows:
\begin{equation}
    \theta_i = 2\arctan{\left( \frac{F_i}{2w} \right)}
    \label{eq:fov}
\end{equation}
where $F_i$ is calculated with the magnification $M$ and number of pixels in the according pixel direction of the \tip{dvs}:
\begin{equation}
    \begin{split}
        F_{\text{x}} &= \frac{346 \text{ [px]}}{M} \\
        F_{\text{y}} &= \frac{260 \text{ [px]}}{M} \\
    \end{split}
    \label{eq:F}
\end{equation}

Table \ref{tab:opening_angles} shows the \tip{aov} $\theta_{\text{x}}$ and $\theta_{\text{y}}$ for the \tip{hddg} and \tip{ivdg} setup. Due to the small values, an assumption of a parallel \tip{fov} is appropriate ($\theta_i=0$). The magnification $M$ in the vicinity of the \tip{pof} can be assumed to be constant, which simplifies the velocity calculation.

\begin{table}[h!]
    \centering
    \renewcommand{\arraystretch}{1.2}
    \begin{tabular}{| l | c | c |}
        \multicolumn{3}{c}{\textbf{Angles of view}} \\
        \hline
        Experiment  & $\theta_{\text{x}}$ [°] & $\theta_{\text{y}}$  [°] \\
        \hline\hline
        \tip{hddg}         & 1.3                     & 1.0                      \\
        \tip{ivdg}         & 5.7                     & 4.3                      \\
        \hline
    \end{tabular}
    \caption{\tip{aov} $\theta_{\text{x}}$ and $\theta_{\text{y}}$ for the \tip{hddg} and \tip{ivdg} experiment in horizontal and vertical direction (x- and y-direction).}
    \label{tab:opening_angles}
\end{table}

\subsubsection{Depth of Field} \label{subs:dof}
The \tip{dof} is approximately given by \eqref{eq:dof}:
\begin{equation} \label{eq:dof}
    \text{DoF}=\frac{2w^2Nc}{f^2}
\end{equation}
\noindent for a given circle of confusion $c$, focal length $f$, $f$-number $N$, and working distance $w$~\autocite{Allen2012-manual-of-photography}. $N$ is the ratio of $f$ to the diameter of the entrance pupil. $c$ is the diameter of the largest image plane circle that is indistinguishable from a point.

Using the \tip{dvs} pixel pitch of 18.5\,um for $c$, $w=50\text{cm}$, $N=4.5$, and $f=300\text{mm}$ for the \tip{hddg} measurements results in $\text{DoF}=0.46\text{mm}$.

A smaller \tip{dof} results in a more pronounced hourglass shape for the accumulated \tip{dvs} events produced by a droplet. 
\eqref{eq:dof} shows that we can minimize the \tip{dof} by using a fast lens (small $N$) and by maximizing the ratio of focal length to working distance ($f/w$).

\subsubsection{Droplet diameter and velocity computation from \tip{dvs} image plane measurement} \label{subs:jaer_measurement}

Given the image plane droplet diameter $d_\text{r}$ (Sec.~\ref{subs:jaer_pics}), the droplet diameter $d$ can be calculated from \eqref{eq:d_DVS}:
\begin{equation}
    d \text{ [mm]} = \frac{d_\text{r}\text{ [px]}}{M \text{ [px/mm]}}.
    \label{eq:d_DVS}
\end{equation}

For droplet velocity measurement, it is useful to use a lens with a long focal length, resulting in a small \tip{aov} $\theta_i$, so that the magnification $M$ at the \tip{pof} can be assumed to be constant. To further mitigate this effect, it is important to measure the velocity as close as possible to the center of the hourglass. The velocity is also measured with the \textit{Speedometer} tool. During playback of a droplet, the velocity is measured by clicking the middle of the droplet at two points surrounding the passage of the droplet through the \tip{pof}, resulting in the \textit{Speedometer} outputting the horizontal and vertical displacement $\Delta{x}_\text{r}$ and $\Delta{y}_\text{r}$ as well as the time difference $\Delta{t}$ and the velocity vector $v_\text{r} \text{ [kpx/s]}$ in px/s. The droplet fall velocity is calculated from \eqref{eq:v_DVS}:
\begin{equation}
    v \text{ [m/s]} = \frac{\sqrt{\left(\Delta x_\text{r} \text{ [px]}\right)^2+\left(\frac{\Delta y_\text{r} \text{ [px]}}{\sin{\alpha}}\right)^2}}{10^3 \cdot \Delta t \text{ [s]} \cdot M \text{ [px/mm]}} \stackrel{(*)}{\approx} \frac{\Delta y_\text{r} \text{ [px]}}{10^3 \cdot \sin{(\alpha)} \cdot \Delta t \text{ [s]} \cdot M \text{ [px/mm]}}     = \frac{v_\text{r} \text{ [kpx/s]}}{\sin(\alpha) \cdot M \text{ [px/mm]}}
    \label{eq:v_DVS}
\end{equation}
\eqref{eq:v_DVS} is simplified to the second form when $\beta$ is small and therefore $(*)$ $\Delta{y}_\text{r} \gg \Delta{x}_\text{r}$. This simplified formula can only be applied if the drop trajectory is parallel to the vertical $yz$-plane (see Fig. \ref{fig:fig1}B: left and Fig. \ref{fig:setup}: bottom left corner) for a correct velocity calculation, otherwise, $\alpha$, which is used for the velocity simulation, would not be constant anymore.
Fig. \ref{fig:fig1}C shows an abstract illustration of a measurement of diameter and velocity from a \tip{dvs} recording, where the black circles show where the actual droplets are. The circles enclose the bottom edge of the ON events (green points) and OFF events (red points). The tracking points of the droplets that we used for the \tip{dvs} velocity estimates are the centers of the black circles. Sec.~\ref{subs:jaer_pics} shows examples of our actual jAER measurements of diameter and velocity.

\subsubsection{Ground truth droplet size measurement}
\label{subs:ground_truth}

Our \tip{hddg} droplets were larger than the 100\,µm droplets which were the focus of \textcite{Kosch2015-hdd-droplet-generator}. The needle frequency is adjusted with a function generator and resulted in droplet diameters between 0.3 and 0.6\,mm, which in our case corresponded to a droplet creation frequency between 60 and 220\,Hz and flow rate 
\num{5.19e-3}\,g/s. At An ISMATEC REGLO Digital peristaltic pump supplied our \tip{hddg} with water at constant flow rate from a tank placed on top of a KERN 440 weighing scale. The scale measured how much water left the tank over time to determine the flow rate. By assuming spherical water drops, their diameter is inferred from their mass, which increases with flow rate but decreases with drop creation frequency. The drop creation frequency describes how many drops per second are produced. According to a water droplet model proposed by \textcite{beard1987new}, a spherical assumption is very accurate for submillimeter drops.

For the \tip{ivdg} experiments (described in \ref{subs:experiments}), the mass of single droplets was determined with a scale (detailed description in \ref{subs:drop_calc}). 
For these droplets, the model by \textcite{beard1987new} predicts a slight average drop deformation due to drag. However, this average deformation is 0.8\% in the case of 2.50\,mm droplets, making them appear to be 2.52\,mm when viewed from roughly 30°. The average deformation thus introduced only a very small systematic \tip{dvs} error. A random error was additionally introduced from the vibration of the 2.5\,mm droplets.

The calculation of the diameter from mass with a spherical assumption is described in \ref{subs:drop_calc} for the \tip{hddg} and \tip{ivdg}. These diameter calculations served as our \tip{gt} values to evaluate the performance of our \tip{dvs} diameter measurements.

\subsubsection{Ground truth droplet diameter from mass} \label{subs:drop_calc}

We used an electronic weighing scale with 0.1mg precision to measure the decrease in mass $\Delta{m}$ of the water tank over a certain period of time $\Delta{t}$. The volumetric flow rate $\dot{V}$ is calculated as \eqref{eq:d_hdd}:
\begin{equation}
    \dot{V} = \frac{\Delta{m}}{\rho \Delta{t}}
\end{equation}
where $\rho$ is the density of water. With the flow rate $\dot{V}$ and the assumption of a sphere, the diameter is calculated as \eqref{eq:d_hdd}:
\begin{equation}
    d = \left( \frac{6 \dot{V}}{\pi f} \right)^{1/3}
    \label{eq:d_hdd}
\end{equation}
where $f$ is the drop creation frequency.

To create larger droplets, we used the \tip{ivdg}. To determine the mass of a single droplet, many droplets were counted and collected in a reservoir positioned on the scale while the total mass $m_{\text{tot}}$ and the number of drops $N$ were recorded. The volume of a single drop is calculated as
\begin{equation}
    V_{\text{drop}} = \frac{m_{\text{tot}}}{\rho N}
\end{equation}
The diameter can then be calculated as
\begin{equation}
    d = \left( \frac{6 V_{\text{drop}}}{\pi } \right)^{1/3}
    \label{eq:d_iv}
\end{equation}
\noindent where the drop is assumed to be a sphere. The diameters calculated from \eqref{eq:d_hdd} and \eqref{eq:d_iv} are used as \tip{gt} values to compare with the diameter measurements of \tip{dvs}.

\subsection{Numerical speed simulation of falling drops} \label{subs:simulation}

A numerical velocity simulation was used to analyze the dynamic behavior of a falling droplet with a diameter up to 2.5\,mm. We used the results of the simulation to determine the terminal speed and the fall height needed to reach any desired final speed, ideally close to the terminal speed. Being close to the terminal speed ensures that \tip{dvs} captures drops with properties similar to real rainfall, and ensures that an uncertainty in height measurement leads to a small deviation from the predicted velocity by simulation.

For simulation, all water drops were assumed to be solid and smooth spheres, which is an eligible approximation especially for drops less than 1\,mm that do not experience any significant deformation according to \textcite{beard1987new} and \textcite{van1997numerical}. Any effect of deformation or vibration due to air drag and turbulence was neglected. Literature values for the air and water properties were used, where the air and water temperatures for 20 and 25°C were interpolated to 22.5°C.

The differential equation for the velocity simulation consists of a drag force, gravity, and acceleration term. The differential equation is numerically solved using the Euler forward method. The relation between drag coefficient and Reynolds number for solid spheres is used, which was fitted to the model of \textcite{yang2015general} with the empirically obtained data of \textcite{brown2003sphere}.

Numerical velocity simulation is used as the velocity \tip{gt} to compare velocity measurements with \tip{dvs}.  To determine the accuracy of the simulation, the terminal velocity of the simulation is compared with the accepted reference data of \textcite{gunn1949terminal}. The results show that they are very close to each other for drops up to 1\,mm. However, for larger drops, the simulation predicts slightly higher velocities; for the 2.5\,mm drops created with the \tip{ivdg}, the simulation predicts 7.9\,m/s whereas \textcite{gunn1949terminal} predict 7.5\,m/s.

\begin{figure}[h!]
    \centering
    \includegraphics[width=\textwidth]{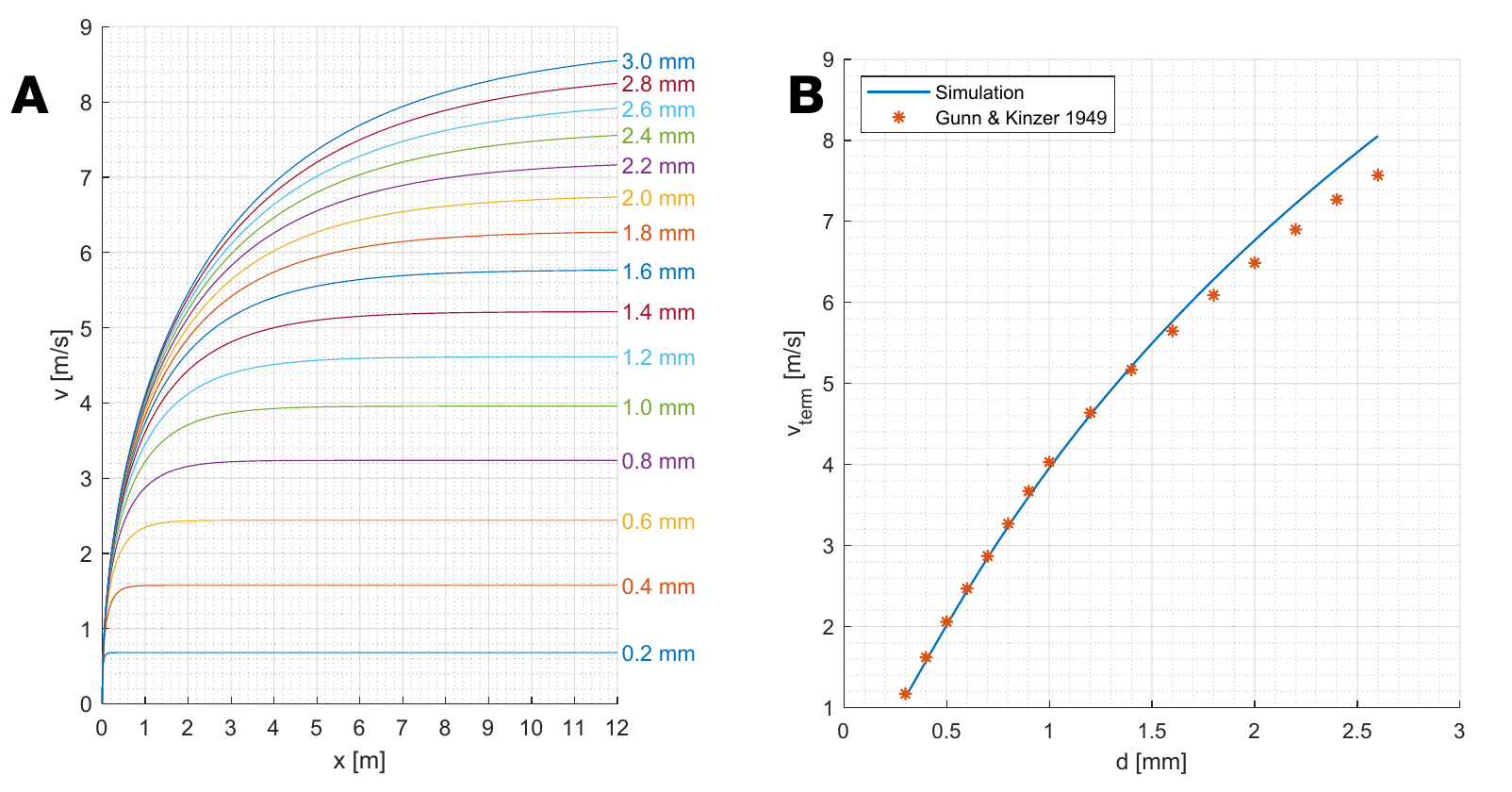}
    \caption{Falling droplet simulation results. \textbf{A:} Velocity simulation for a fall height up to 12\,m and for different drop diameters (written on the right side); vertical displacement $y$ vs. velocity $v$. Literature values for the density of air $\rho_{\text{air}}$, kinematic viscosity of air $\nu_{\text{air}}$ and density of water $\rho_{\text{water}}$ were used for a temperature of 20°C.
    \textbf{B:} Comparison of the terminal simulated terminal velocity to the data of \textcite{gunn1949terminal}; diameter $d$ vs. terminal velocity $v_{\text{term}}$. Literature values for the density of air $\rho_{\text{air}}$, kinematic viscosity of air $\nu_{\text{air}}$ and density of water $\rho_{\text{water}}$ were used for a temperature of 20°C.}
    \label{fig:simulation_plot}
\end{figure}

Fig. \ref{fig:simulation_plot}A plots the model for different drop diameters. For larger droplets, the fall height must be higher to reach the terminal velocity. Moreover, the terminal velocity for larger drops is larger than for smaller ones.
Fig. \ref{fig:simulation_plot}B compares the simulated velocity and the measured data from \textcite{gunn1949terminal}. The simulation starts to differ from the data for droplets with a diameter above 1.5\,mm.

According to our simulation, a fall height of 2\,m is sufficient for drops with a diameter of up to 0.6\,mm to reach 99\% of the terminal velocity, while for 2.5\,mm drops, a fall height of 10\,m is necessary to reach 97\% of the terminal velocity (see Fig. \ref{fig:simulation_plot}). Thus, a fall height of 2\,m was used for the \tip{hddg} experiment and 10\,m was used for the \tip{ivdg} experiment.

\subsubsection{Details of droplet speed simulation}
This following describes the details of the model used to simulate the speed of a falling water droplet.  The goal is to find the velocity $v$ of the drop as a function of the vertical distance $y$ the drop has traveled from its initial position for any given diameter $d$. This model allows us to determine the terminal speed and the required fall height to reach a certain fraction of the terminal speed $v_\text{term}$, ideally close to the terminal speed.
Our model is based on \textcite{yang2015general}.

The drag force $F_\text{D}$ acts on a falling water drop that eventually reaches equilibrium at terminal velocity $v_\text{term}$ with the gravitational force $mg$. The drag force is defined as \eqref{drag_force}:
\begin{equation}
    F_{\text{D}} = \frac{1}{2}\rho_{\text{air}}c_{\text{D}}(Re)Av^{2} = k(Re)v^{2}
    \label{drag_force}
\end{equation}
\noindent where $c_\text{D}$ is the drag coefficient depending on the Reynolds number $Re$, $\rho_{\text{air}}$ is the density of air, $A$ is the area facing the fluid (for spheres: $A=\pi{}(\frac{d}{2})^{2}$) and $k=\frac{1}{2}\rho_{\text{air}}c_{\text{D}}A$.  
The equation of motion can be derived as \eqref{eq_motion}:
\begin{equation}
    ma = mg - F_{\text{D}} = mg-k(Re)v^{2}
    \label{eq_motion}
\end{equation}
\noindent  where $a$ is the acceleration, $v$ is the velocity, and $g$ is the gravitational acceleration.
\eqref{eq_motion} can be rewritten as a differential equation:
\begin{equation}
    m\ddot{y} = mg-k(Re)\dot{y}^{2}
    \label{eq_diff}
\end{equation}
\noindent where $y$ describes the vertical displacement of the droplet from the droplet generator.
As mentioned above, the drag coefficient $c_{\text{D}}$ depends on the Reynolds number $Re$. The curve fit for the drag coefficient derived by \textcite{yang2015general} is used for the simulation, which is expressed as \eqref{drag_coeff_fit}:
\begin{equation}
    \begin{split}
        x &= \frac{\ln{\left(1+Re\right)}}{10} \\
        \alpha &= \left[1-\exp{\left(-3.24x^2 + 8x^4 - 6.5x^5\right)}\right]\cdot\frac{\pi}{2} \\
        c_{\text{D}} &= \frac{24}{Re}\cdot\left(1 + \frac{3}{16}Re\right)^{0.635}\cdot\cos^3{\alpha} + 0.468\cdot\sin^2{\alpha}
    \end{split}
    \label{drag_coeff_fit}
\end{equation}
where the Reynolds number $Re$ is defined as
\begin{equation}
    Re = \frac{\rho vL}{\mu} = \frac{vL}{\nu},
\end{equation}
where $\rho$ is the density of the fluid, $v$ is the flow velocity, $L$ is the characteristic length (in this case $L=d$), $\mu$ is the dynamic viscosity of the fluid, and $\nu$ is the kinematic viscosity of the fluid.  \eqref{drag_coeff_fit} is a very good approximation for Reynolds numbers $Re < \num{2e5}$, which water drops never exceed, according to \textcite{gunn1949terminal}. 


\MATLAB code to compute discrete time updates of these equations is available from our cloud drive (see Sec.~\ref{sec:data_availability}).





\printbibliography


\end{document}